\documentclass[lettersize,journal]{IEEEtran}
\usepackage{amsmath,amsfonts}
\usepackage{algorithmic}
\usepackage{algorithm}
\usepackage{array}
\usepackage[caption=false,font=normalsize,labelfont=sf,textfont=sf]{subfig}
\usepackage{textcomp}
\usepackage{stfloats}
\usepackage{url}
\usepackage{verbatim}
\usepackage{graphicx}
\usepackage{cite}
\usepackage{color}

\begin{document}

\title{Exploiting Structured Sparsity with Low Complexity Sparse Bayesian Learning for RIS-assisted MIMO Channel Estimation}

\author{Weijie Li, Zihuai Lin,~\IEEEmembership{Senior Member,~IEEE}, Qinghua Guo,~\IEEEmembership{Senior Member,~IEEE},\\ Branka Vucetic,~\IEEEmembership{Life Fellow,~IEEE}
        % <-this % stops a space
%\thanks{This paper was produced by the IEEE Publication Technology Group. They are in Piscataway, NJ.}% <-this % stops a space
%\thanks{Manuscript received April 19, 2021; revised August 16, 2021.}}
\thanks{Weijie Li, Zihuai Lin and Branka Vucetic are with the School of Electrical and Information Engineering, University of Sydney, Sydney, NSW 2006, Australia (e-mail: weli0332@uni.sydney.edu.au; zihuai.lin@sydney.edu.au; branka.vucetic@sydney.edu.au).

Qinghua Guo is with the School of Electrical, Computer and Telecommunications Engineering, University of Wollongong, Wollongong, NSW 2522, Australia (e-mail: qguo@uow.edu.au) }}

% \markboth{Journal of \LaTeX\ Class Files,~Vol., No., June~2023}%

\IEEEpubid{}

\maketitle

\begin{abstract}
As an emerging communication auxiliary technology, reconfigurable intelligent surface (RIS) is expected to play a significant role in the upcoming 6G networks. Due to its total reflection characteristics, it is challenging to implement conventional channel estimation algorithms. This work focuses on RIS-assisted MIMO communications. Although many algorithms have been proposed to address this issue, there are still ample opportunities for improvement in terms of estimation accuracy, complexity, and applicability. To fully exploit the structured sparsity of the multiple-input-multiple-output (MIMO) channels, we propose a new channel estimation algorithm called unitary approximate message passing sparse Bayesian learning with partial common support identification (UAMPSBL-PCI). Thanks to the mechanism of PCI and the use of UAMP, the proposed algorithm has a lower complexity while delivering enhanced performance relative to existing channel estimation algorithms. Extensive simulations demonstrate its excellent performance in various environments.
\end{abstract}


\section{Introduction}
\IEEEPARstart{D}{ata} traffic has risen in recent decades due to the quickly growing mobile equipment and software industries, which has greatly raised the demand for effective communication networks. In order to further enhance the system capacity and transmission rate, attention has turned to high-frequency bandwidths, such as the millimeter wave (mmWave) and Terahertz (THz) frequency bands, in both current 5G networks and upcoming 6G networks \cite{xu2021THzUAV,AmBClikun,5G_Yang1,5G_Yang2,network_capacity,URLLC}. Recently, a promising new technology called reconfigurable intelligent surface (RIS) has emerged as an innovative solution \cite{RN112,RN114,commLettlikun,chu2022IoT_RIS,WirelessCommLettlikun}.  Typically, RIS is a uniform planar array composed of electromagnetic materials on which a large number of passive reflective elements are deployed \cite{RN5,RN110,RN111}. Each of these reconfigurable reflective units can independently induce a phase shift in the incoming signal according to the surrounding environment \cite{RN6}.

RIS has numerous advantages over current wireless technologies. First, a typical RIS is composed of passive reflecting units with no active components or RF chain. This design significantly reduces energy consumption and layout costs. Second, RIS is capable of operating in full-duplex mode, resulting in superior spectral efficiency. Additionally, RIS can serve as compatible supporting technology for existing cellular systems, facilitating the use of both direct and reflection channels for signal transmission, thereby enhancing the received power. Moreover, the RIS system's powerful anti-scatter capability increases signal coverage and propagation distance, resolving effectively signal blind areas and $``$dead zones$"$.

Despite the fact that RIS has several advantages over existing wireless communication technologies, it still confronts a number of problems that must be overcome before it can be extensively commercialized \cite{RN7,RN8,RN104}.%,RN107,RN108}. 
Among them, the most apparent and significant one is channel estimation \cite{RN7}. On the one hand, the large-scale use of reflecting units in RIS leads to the formation of a vast number of reflecting channels, which exponentially increases the number of channel parameters that require estimation. This greatly increases the complexity and difficulty of channel estimation. On the other hand, since the deployed reflecting units are passive and lack signal transmission capability, traditional channel estimation methods that rely on transmitting channel pilots cannot be used. Furthermore, the highly variable layout conditions and system setup of the RIS system create varying demands on channel estimation requirements and constraints, which need to be specified and adjusted accordingly \cite{RN8}.

Many attempts have been made to solve the cascaded channel estimation issue formed by the fully passive RIS systems \cite{RN9,RN10,RN11,RN12,RN13,RN14}. In \cite{RN9}, the traditional least square (LS) algorithm was enhanced by incorporating a new ON/OFF switching reflection mechanism. To be specific, it is operated by turning off all reflection units on the RIS panel first, then turning them on one by one to estimate the cascaded channel information generated by each reflection unit, and ultimately integrating them into the complete channel state information (CSI). The compressed sensing (CS) algorithm was adopted in \cite{RN10} to achieve efficient and accurate channel estimation results of the target system. This algorithm takes full advantage of the sparse and low-rank properties of the RIS-assisted system channel in the high-frequency band, simplifying the complex channel estimation problem to a more straightforward sparse channel recovery problem. Channel decomposition is another promising technique that has been used. \cite{RN11} applied matrix factorisation to estimate the low-dimensional channels obtained by decomposition, which requires less pilot overhead and computational effort. Recently, deep learning-based channel estimation algorithms have been applied to this system. Due to the fact that the channel estimation problem can be approximated as a nonlinear mapping from the received signal to the channel information, the powerful data processing capability of deep learning can significantly reduce the training overhead and time. \cite{RN12} and \cite{RN13} have proposed deep learning-based channel estimation schemes using convolutional neural networks (CNN) and deep neural networks (DNN), respectively. Federated learning (FL) \cite{leng2022FL,Nguyen2022FL} was also considered for this problem \cite{RN14}.

In this paper, we consider a fully passive RIS-assisted mmWave MIMO communication system \cite{5G_sina,MIMO_capacity,Chen2021MIMO}. The recently introduced methodologies for channel estimation demonstrate superior performance and reduced training duration when compared to conventional channel estimation algorithms. This is achieved through the utilization of angular domain sparsity inherent in the mmWave channel. However, there are two main concerns that still need to be addressed. The first one is the complexity. Commonly used sparse channel estimation algorithms typically achieve great performance at the cost of high complexity \cite{RN8}, which hinders their applications in real communication systems. The second one is the applicability. In the channel estimation for the RIS-assisted communication system, a common assumption is that all users share some common paths \cite{RN10}, but this is rare in real systems. We propose a novel channel estimation algorithm based on UAMP-SBL to tackle the identified issues. The main contributions of this paper are given below.

\begin{itemize}
\item {We build a sparse system model based on the dual-structure sparsity of the RIS angular cascaded channel and extend it from a simple system where all users share the common channels (denoted as Scenario 1) to a more general system where only subsets of users share common channels (denoted as Scenario 2). In this more general system, channel estimation becomes more challenging, and the existing algorithms do not perform well.}
\item{We propose a new channel estimation algorithm called unitary approximate message passing sparse Bayesian learning with partial common support identification (UAMPSBL-PCI), which delivers better performance but with lower complexity compared with the existing channel estimation techniques. This algorithm not only extends the single measurement vector (SMV) algorithm to the multiple measurement vector (MMV) algorithm, but also leverages the sparse prior of the communication channels.}
\item{As the channel characteristics of the simple system in Scenario 1 and the more general system in Scenario 2 are distinct, we devise algorithms to accommodate their differences.}
\item{We evaluate the estimation performance of the proposed algorithm under different system parameter settings and demonstrate its wide applicability through extensive simulations.}
\end{itemize}

The rest of this paper is organized as follows. In Section II, the model of the RIS-assisted communications system and corresponding sparse channel structure are presented. The channel estimation problem is also formulated. In Section III, UAMP-SBL based channel estimators for both simple and general scenarios are proposed. In Section IV, simulation results are provided to demonstrate the superiority of the proposed algorithm. Finally, this paper is concluded in Section V. 

Notations: Lower case a, lower case boldface $\mathbf{a}$ and upper case boldface A denote scalar, vector and matrix, respectively. $\mathbf{0}$ and $\mathbf{1}$ are properly sized vectors with all elements are $0$ and $1$. $\mathbf{(\cdot)}^{T}$, and $\mathbf{(\cdot)}^{H}$ denote the transpose and conjugate transpose of a matrix. The notation $\mathbf{a}\sim\mathcal{C}\mathcal{N}(\mu,\Sigma)$ means $\mathbf{a}$ follows the complex Gaussian distribution with mean $\mu$ and covariance $\Sigma$. $\mathbf{a}_{i}$ is the $i$-th element of $\mathbf{a}$. diag$(\mathbf{a})$ denotes a diagonal matrix with the diagonal element $\mathbf{a}$. $\otimes$ denotes the Kronecker product.$\left\| \ \right\|$ is the $l_{2}$ norm.
${Ga\left( \gamma \middle| \epsilon,\eta \right)}$ is a Gamma distribution.

\section{System model and Problem formulation}
In this section, we first introduce the system model of RIS-assisted mmWave MIMO uplink transmission. Then, the channel structure in the angular domain for both Scenarios 1 and 2 are presented. At last, we formulate the channel estimation problem that fully exploits the dual-structure sparsity of the cascaded channel.

\subsection{RIS-assisted mmWave MIMO Uplink Transmission}

We consider a RIS-assisted mmWave MIMO communication system shown in Fig. 1. The base station (BS) is equipped with an $M$-antenna uniform planar array (UPA). A fully passive RIS composed of $N$ elements is positioned between the BS and the user zone. There are $J$ users in the user zone, and each user is a single-antenna terminal. The direct line-of-sight (LoS) communication channels between the BS and users are assumed to be blocked by obstacles. Thus, only indirect none-line-of-sight (NLoS) channels through RIS are considered in this system. The channel between  the BS and RIS is denoted as 
$\mathbf{H}_{BR}$, and the channel between RIS and the $j$-th user is denoted as 
$\mathbf{h}_{j}$. Following \cite{RN15}, the Saleh-Valenzuela channel model is adopted, which emulates the signal propagation effects by characterizing the time delay and fading amplitude of each path within the multipath channel. The BS-RIS channel  $\mathbf{H}_{BR} \in \mathbb{C}^{M \times N}$ is given as 
\begin{equation}
\mathbf{H}_{BR} = \sqrt{\frac{MN}{P_{BR}}}{\sum\limits_{\rho = 1}^{P_{BR}}{\alpha_{\rho}^{BR}~\mathbf{b}\left( {\theta_{\rho}^{B},\phi_{\rho}^{B}} \right)\mathbf{r}^{T}\left( {\theta_{\rho}^{R_{1}},\phi_{\rho}^{R_{1}}} \right),~~}}
\end{equation}
where $P_{BR}$ is the number of paths between the BS and RIS while $\alpha_{\rho}^{BR}$ denotes the channel gain of the $\rho$-th path, $\theta_{\rho}^{B}$ and $\theta_{\rho}^{R_{1}}$ are the azimuth angle of the $\rho$-th path at the BS and RIS respectively, $\phi_{\rho}^{B}$ and $\phi_{\rho}^{R_{1}}$ are the elevation angle of the $\rho$-th path at the BS and RIS respectively, and $\mathbf{b}\left( {\theta,\phi} \right)$ and $\mathbf{r}\left( {\theta,\phi} \right)$ are the normalised steering vector of the RIS and BS respectively. As shown in \cite{RN15}, the normalised steering vector of an $L$-antenna UPA with $L_{r}$ rows and $L_{c}$ columns can be written as 
\begin{equation}
\mathbf{s}\left( {\theta,\phi} \right) = \frac{1}{\sqrt{L}}~\mathbf{s}_{\mathbf{r}} \otimes \mathbf{s}_{\mathbf{c}},
\end{equation}
where
\begin{equation}
\mathbf{s}_{\mathbf{r}} = e^{ -j2\pi dsin{(\theta)}{\cos{(\phi)}} \times \frac{\mathbf{l}_\mathbf{r}}{\lambda}}, \ \mathbf{s}_{\mathbf{c}} = e^{ -j2\pi dsin{(\phi)} \times \frac{\mathbf{l}_\mathbf{c}}{\lambda}}
\end{equation}
with ${\mathbf{l}_\mathbf{r}} = \left\lbrack 0,1,2,\ldots,\left( L_{r} - 1 \right) \right\rbrack$, ${\mathbf{l}_\mathbf{c}} = \left\lbrack 0,1,2,\ldots,\left( L_{c} - 1 \right) \right\rbrack$. Typically, the UPA has a  half-wavelength antenna spacing ($d = \frac{\lambda}{2}$). By substituting (3) into (2), we can finally get the expression for the normalized steering vector of the UPA as
\begin{equation}
\mathbf{s}\left( {\theta,\phi} \right) = \frac{1}{\sqrt{L}}{\left\lbrack e^{ -j2\pi dsin{(\theta)}{\cos{(\phi)}} \times \frac{\mathbf{l}_\mathbf{r}}{\lambda}} \right\rbrack \otimes \left\lbrack e^{ -j2\pi dsin{(\phi)} \times \frac{\mathbf{l}_\mathbf{c}}{\lambda}} \right\rbrack}.
\end{equation}

\begin{figure}[!t]
\centering
\includegraphics[width=3.5in]{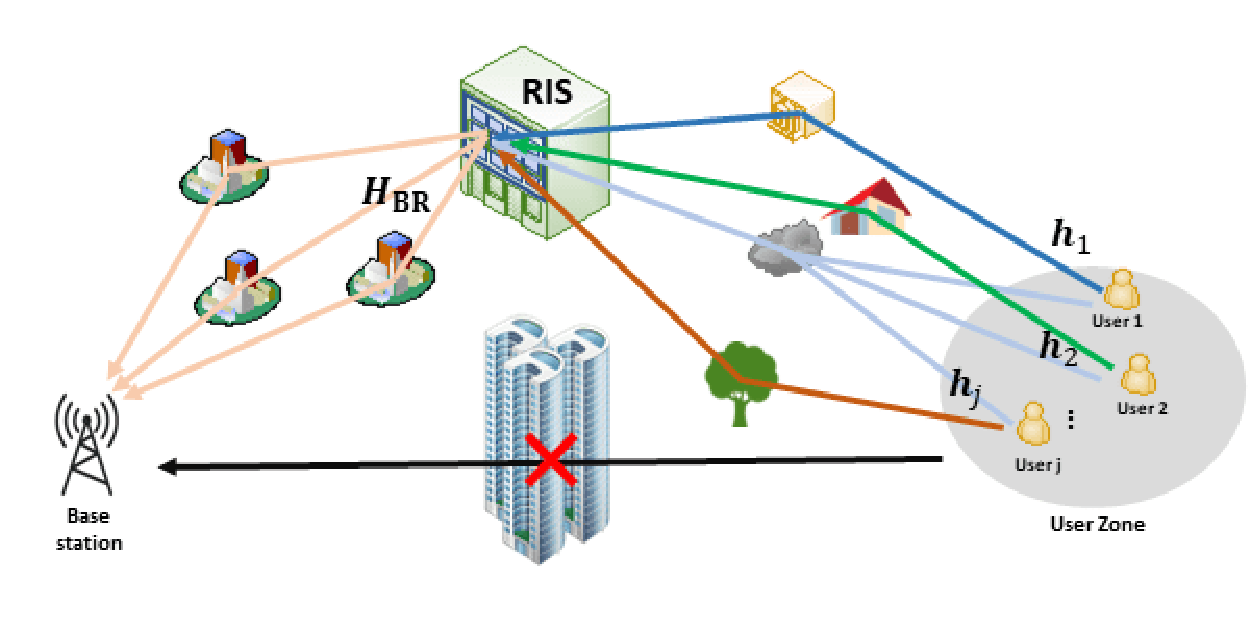}
\caption{Scenario 1: simple system where all users share the common channels.}
\label{fig_1}
\end{figure}

The RIS-User channel $\mathbf{h}_{j} \in \mathbb{C}^{N \times 1},j = \{1,2,\ldots,J\}$ can also be modelled as 
\begin{equation}
\mathbf{h}_{j} = \sqrt{\frac{N}{P_{j}}}{\sum\limits_{\varsigma = 1}^{P_{j}}{\alpha_{\varsigma}^{j}~\mathbf{r}\left( \theta_{\varsigma}^{R_{2}},\phi_{\varsigma}^{R_{2}} \right),~}}
\end{equation}
where $P_j$ is the number of paths between the RIS and $j$-th user while $\alpha_{\varsigma}^{j}$ denotes the channel gain of the $\varsigma$-th path, and $\theta_{\varsigma}^{R_{2}}$ and $\phi_{\varsigma}^{R_{2}}$ are the respective azimuth and elevation  angle of the $\varsigma$-th path at the RIS. It is noted that since RIS receives the signal from the transmitter and reflects it to the receiver, it has both angle-of-arrival (AoA) and angle-of-departure (AoD). Here, $R_{1}$ represents AoD and $R_{2}$ represents AoA.

Then, we can finally obtain the cascaded communication channel $
\mathbf{H}_{j} \in \mathbb{C}^{M \times N}$ from the $j$-th user to the BS, which can be formulated as
\begin{equation}
\mathbf{H}_{j} = \mathbf{H}_{BR}~\text{diag}\left( \mathbf{h}_{j} \right).
\end{equation}

Since the BS equipped with UPA antenna has better signal processing capability than the single-antenna user terminal, we consider an uplink channel estimation scenario in this work. As the LoS communication channel is blocked by obstacles, the direct channel impact will be ignored.

Assume that in a total of $T$ time slots, pilot symbols are transmitted from the users to the BS through RIS. The received signal from the $j$-th user at the $t$-th time slot, which is denoted by $
\mathbf{y}_{j}^{t} \in \mathbb{C}^{M \times 1}$ , can be written as 
\begin{equation}
\mathbf{y}_{j}^{t} = \mathbf{H}_{BR}~\text{diag}\left( \boldsymbol{\omega}_{t} \right)\mathbf{h}_{j}x_{j}^{t} + \mathbf{w}_{j}^{t},
\end{equation}
where  $\boldsymbol{\omega}_{t} \in \mathbb{C}^{M \times 1}$ stands for the phases of the RIS at the $t$-th time slot, which is the same for all users, and $x_{j}^{t}$ denotes the pilot symbol. For simlicity, $x_{j}^{t}$ is set to 1 in this paper. $\mathbf{w}_{j}^{t}\in{\mathbb{C}}^{M \times 1} \sim \mathcal{C} \mathcal{N}(0,{\sigma}^{2}\mathbf{I}_{M})$ is the Additive white Gaussian noise (AWGN). Due to $\text{diag}\left( \mathbf{a} \right)\mathbf{b} = \text{diag}\left( \mathbf{b} \right)\mathbf{a}$, (7) can be rewritten as
\begin{equation}
\mathbf{y}_{j}^{t} = \mathbf{H}_{j}\boldsymbol{\omega}_{t} + \mathbf{w}_{j}^{t}.
\end{equation}
Then the received signal for the T time slots $\mathbf{Y}_{j} \in \mathbb{C}^{M \times T}$ at the BS can be represented as
\begin{equation}
\mathbf{Y}_{j} = \mathbf{H}_{j}\mathbf{\Omega} + \mathbf{W}_{\mathbf{j}},
\end{equation}
where $\mathbf{\Omega} \in \mathbb{C}^{N \times T}$ includes the phase vector set at the RIS, which is known at the channel estimation stage, and $\mathbf{W} \in \mathbb{C}^{M \times T}$ is the AWGN matrix.

\subsection{Channel Structure for Scenario 1}
To fully exploit the structure sparsity, singular value decomposition (SVD) can be performed on $\mathbf{H}_{j}$, in which case $\mathbf{H}_{j}$ can be described as
\begin{equation}
\mathbf{H}_{j} = \mathbf{U}\Breve{\mathbf{H}}_{j}\mathbf{V}^{H},
\end{equation}
where  $\Breve{\mathbf{H}}_{j}$ is the angular cascaded channel which shows great sparsity, $\mathbf{U}$ and $\mathbf{V}$ are dictionary unitary matrices \cite{RN16}. Typically, these dictionary matrices are orthogonal matrices. 

 By substituting (1), (5), and (6) into (10), we can rewrite the angular cascaded channel $\Breve{\mathbf{H}}_{j}\in \mathbb{C}^{M \times N}$ as
\begin{equation}
\begin{aligned}
\Breve{\mathbf{H}}_{j} & = \sqrt{\frac{MN^2}{P_{BR}P_{j}}}{\sum\limits_{\rho = 1}^{P_{BR}}{\sum\limits_{\varsigma = 1}^{P_{j}}{\alpha_{\rho}^{BR}\alpha_{\varsigma}^{j}}}}\mathbf{U}^{H}\mathbf{b}\left( {\theta_{\rho}^{B},\phi_{\rho}^{B}} \right) \\ 
& \quad \ \mathbf{r}^{H}\left( {\theta_{\rho}^{R_{1}} + \theta_{\varsigma}^{R_{2}},\phi_{\rho}^{R_{1}} + 
\phi_{\varsigma}^{R_{2}}} \right)\mathbf{V} \\ 
 & \small{= C_{1}\mathbf{~}{\sum\limits_{\rho = 1}^{P_{BR}}{\sum\limits_{\varsigma = 1}^{P_{j}}C_{\rho,\varsigma}}}\Breve{\mathbf{b}}\left( {\theta_{\rho}^{B},\phi_{\rho}^{B}} \right){\Breve{\mathbf{r}}}^{H}\left( {\theta_{\rho}^{R_{1}} + \theta_{\varsigma}^{R_{2}},\phi_{\rho}^{R_{1}} + \phi_{\varsigma}^{R_{2}}} \right)},
\end{aligned}
\end{equation}
where $C_{1}$ and $C_{\rho,\varsigma}$ are constant, both $\Breve{\mathbf{b}}\in \mathbb{C}^{M \times 1}$ and $\Breve{\mathbf{r}}^{H}\in \mathbb{C}^{1 \times N}$ have only one non-zero element for each set of $(\rho,\varsigma)$. Thus, $\Breve{\mathbf{H}_{j}}$ has a total of ${P_{BR}P_{j}}$ non-zero elements and their sparse rate can be calculated as ${{P}_{BR}}{{P}_{j}}/MN$. In addition, since the row and column index of the non-zero elements is determined by $\Breve{\mathbf{b}}$ and $\Breve{\mathbf{r}}^{H}$ respectively, we can further exploit the correlation at these non-zero positions.

As shown in Fig. 1, since the locations of the BS and RIS are fixed, the cascaded channel of each user shares the common BS-RIS channel $\mathbf{H}_{BR}$, resulting in the direction $\left( {\theta_{\rho}^{B},\phi_{\rho}^{B}} \right)$ of the array steering vector $\mathbf{b}$ being the same for all users. Thus, from the perspective of the channel matrices, the non-zero elements of each user share the complete common ${P}_{BR}$  row indices. Similarly, because all users are assumed to be clustered in a user zone, and their positions are relatively close to each other, the RIS-User channel of each user $\mathbf{h}_{j}, j \in \{1,2,\ldots,J\}$ will share partial common paths. These partial common paths have the same array steering vector ${\mathbf{r}}\left( {\theta_{\rho}^{R_{1}} + \theta_{\varsigma}^{R_{2}},\phi_{\rho}^{R_{1}} + \phi_{\varsigma}^{R_{2}}} \right)$ at RIS, while other non-common paths do not. So the non-zero elements produced by these partial common paths lie on the same columns of channel matrices, and others are randomly distributed according to their unique array steering vector. Here, we introduce a new variable ${{P}_{c}}$ that represents the number of partial common paths. 

\begin{figure}[!t]
%\centering
\hspace{-1.5mm}\includegraphics[width=3.7in]{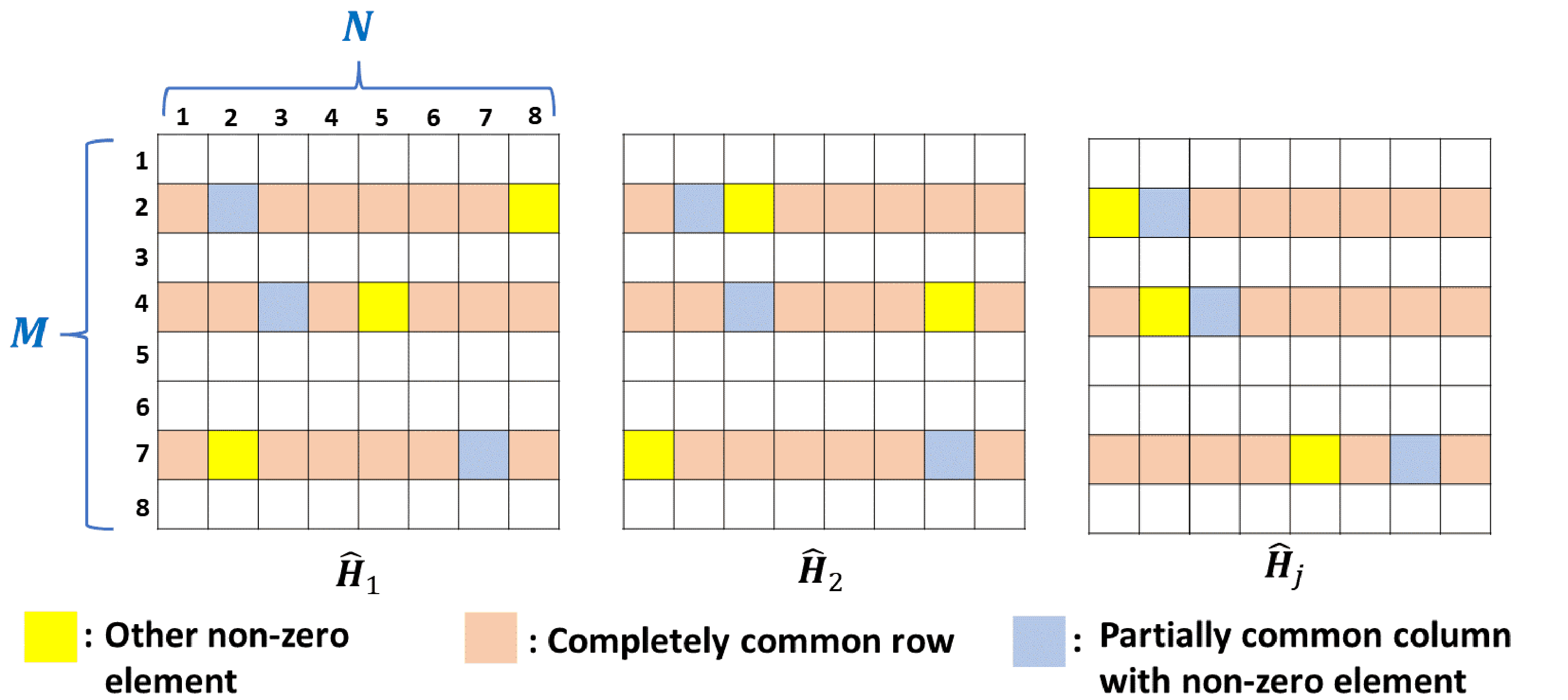}
\caption{The distribution of non-zero elements in the cascaded channel for Scenario 1.}
\label{fig_2}
\end{figure}

The schematic diagram of the distribution of non-zero elements in the cascaded channel is presented in Fig. 2. In sum, for the set of angular cascaded channel  $\Breve{\mathbf{H}}_{j} \in \mathbb{C}^{N \times 1}, j \in \{1,2,\ldots,J\}$, each of them has a total of ${{P}_{BR}}{{P}_{j}}$ non-zero elements. These non-zero elements fully share ${{P}_{BR}}$ common rows(e.g. the second, fourth and seventh row in Fig. 2), and ${{P}_{c}}$ partial common columns in each common row (e.g. the second column of the second row, the third column of the fourth row, and the seventh column of the seventh row in Fig. 2). Other non-zero elements will be randomly distributed.

\subsection{Channel Structure for Scenario 2}

In Scenario 1 as stated above, due to the close distance of each user, we can assume that all users share partial common RIS-User channels. However, this kind of system setting is rare in reality. In the following, we will further extend this simple system to a more general system where only subsets of users share the common channels.

As shown in Fig. 3, end users are randomly distributed among $K$ subsets, and the number of users included in each subset is also random, which is denoted as $J_k$. At this point, all users still fully share the BS-RIS channel $\mathbf{H}_{BR}$ due to the fixed positions of the base station and the RIS, while three situations may occur in the RIS-User channel $
\mathbf{h}_{k,j} \in \mathbb{C}^{N \times 1},k \in \left\{ {1,2,\ldots,K} \right\},j \in \left\{ 1,2,\ldots,J_k \right\}$. First, in the same subset, we still consider it to have the same channel characteristics as the simple system where users share partial common paths, e.g. $\mathbf{h}_{1,1}$ and $\mathbf{h}_{1,4}$. Second, although some users are in different subsets, they still have partial common paths when transmitting information due to their relatively close locations, e.g. $\mathbf{h}_{1,4}$ and $\mathbf{h}_{2,1}$. Last but not least, when users are in different subsets and relatively far away, we consider their transmission paths completely independent, and there is no common path e.g. $\mathbf{h}_{1,1}$ and $\mathbf{h}_{2,3}$.

According to (11), the variation in the direction $(\theta,\phi)$ of the array steering vector at the RIS, caused by different principles of sharing RIS-User channels among users, directly affects the distribution of non-zero elements in the sparse angular cascaded channel. The new schematic diagram of the distribution of the non-zero elements in this more general system is presented in Fig. 4. It is clear to see that there is no common column support for all users in each common row, which further increases the difficulty and complexity of channel estimation.

\subsection{Problem Formulation}
Due to the high frequency, short wavelength, limited diffraction effects, and significant path loss encountered during propagation, the  mmWave channel typically exhibits prominent sparse characteristics. The traditional channel estimation problem can usually be modelled as a sparse matrix recovery problem. In order to fully use the significant  sparsity of the cascaded channel, (10) is merged into (9). The received signal can be further written as  
\begin{equation}
\mathbf{Y}_{j} = \mathbf{U}\Breve{\mathbf{H}}_{j}{\mathbf{V}^{H}}\mathbf{\Omega} + \mathbf{W}_{\mathbf{j}}.
\end{equation}

\begin{figure}[!t]
\centering
\includegraphics[width=3.8in]{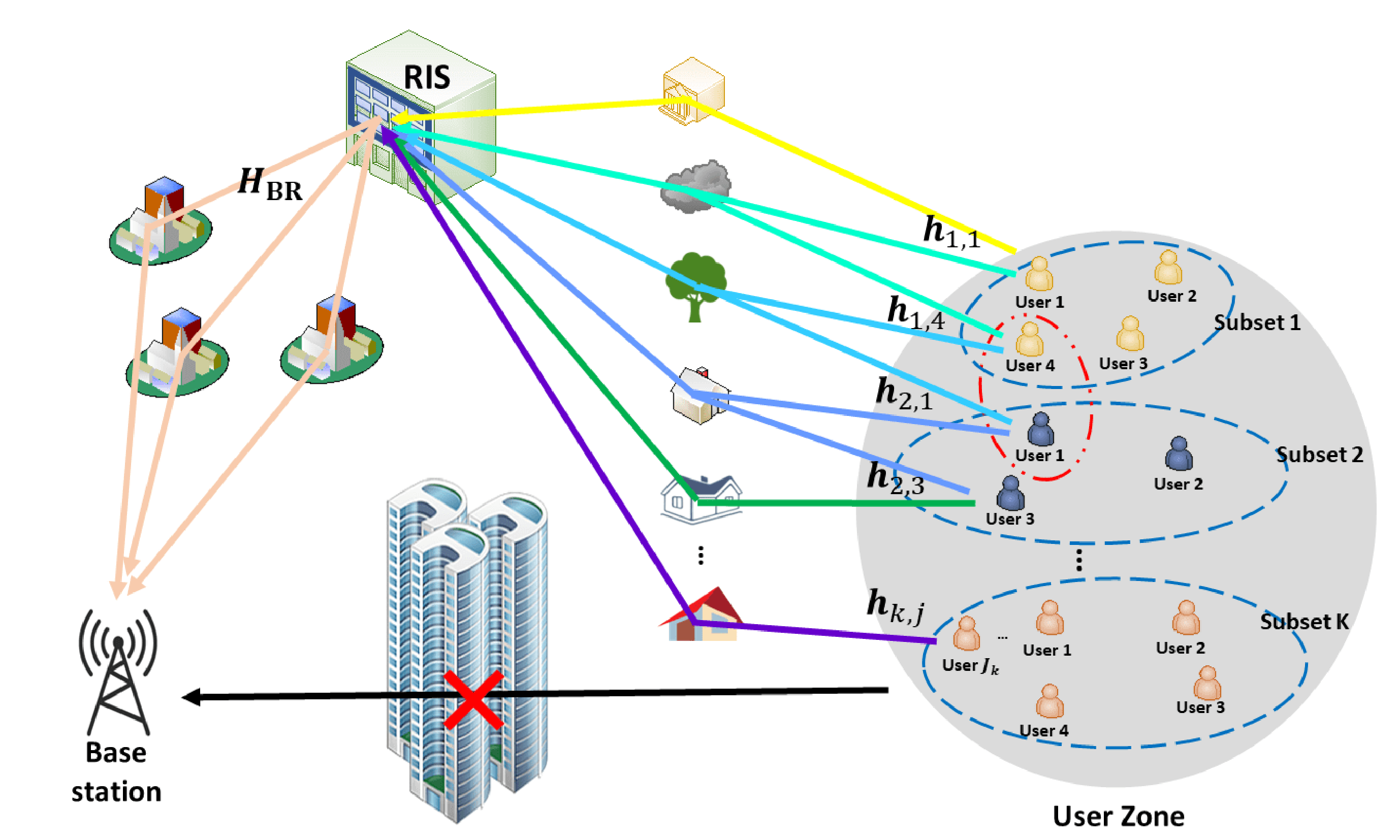}
\caption{Scenario 2: More general system where subsets of users share the common channels.}
\label{fig_3}
\end{figure} 

\begin{figure}[!t]
\centering
\includegraphics[width=3.5in]{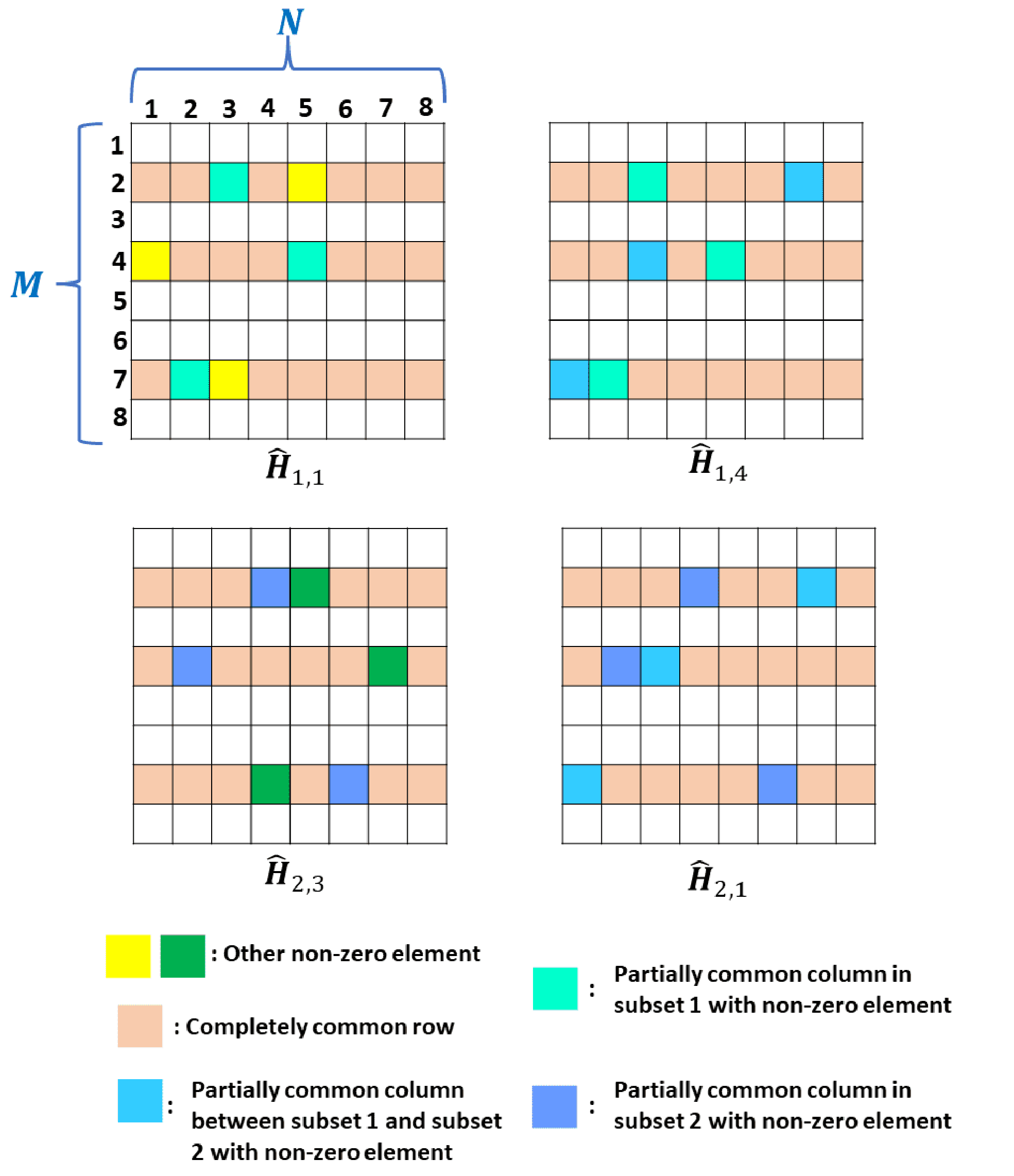}
\caption{The distribution of non-zero elements in the cascaded channel for Scenario 2.}
\label{fig_4}
\end{figure}
To formulate this channel estimation problem as a sparse matrix recovery problem, (12) needs to be transformed into a typical CS model:
\begin{equation}
\Breve{\mathbf{Y}}_{j} = \Breve{\mathbf{\Omega}}{\Breve{\mathbf{H}}_{j}}^{H} + \Breve{\mathbf{W}}_{\mathbf{j}},
\end{equation}
where measurement matrix $\small{
\Breve{\mathbf{Y}}_{j} \in \mathbb{C}^{T \times M} = ( {\mathbf{U}^{H}\mathbf{Y}_{j} )^{H}}}$, sensing matrix $\small{
\Breve{\mathbf{\Omega}} \in \mathbb{C}^{T \times N} = ( {\mathbf{V}}^{H}\mathbf{\Omega} )^{H}}$, and AWGN matrix $\small{
\Breve{\mathbf{W}}_{\mathbf{j}} \in \mathbb{C}^{T \times M} = ( {\mathbf{U}}^{H}\mathbf{W}_{\mathbf{j}} )^{H}}$.

Here, we aim to estimate the cascaded channel matrix $\Breve{\mathbf{H}}_{j}^{H}$ based on the known received signal matrix $\Breve{\mathbf{Y}}_{j}$ and RIS reflecting matrix $\Breve{\mathbf{\Omega}}$ for both Scenarios 1 and 2. By exploiting the structured sparsity, the corresponding $\Breve{\mathbf{H}}_{j}^{H}$ for different users can be estimated jointly, which greatly enhances the estimation performance. It should be noted that due to the absence of common column support for all users in Scenario 2, its channel estimation has greater challenges. In this paper, we propose to use UAMP-SBL based channel estimators to tackle them.

\section{UAMP-SBL Based Channel Estimators Exploiting Structured Sparsity } 
In this section, we first briefly introduce the UAMP-SBL algorithm \cite{RN16}, based on which we design the UAMPSBL-PCI channel estimators for Scenarios 1 and 2, where the dual-structure sparsity of the cascaded channel is fully exploited. The complexity of the proposed channel estimator is also analyzed.

\subsection{UAMP-SBL}

%For a typical CS-modelled single measurement vector (SMV) sparse matrix recovery problem 
The SBL algorithm \cite{RN20} can be used to recover a length-$N$ sparse vector $\mathbf{x}$ with the following model
\begin{equation}
\mathbf{y} = \mathbf{S}\mathbf{x} + \mathbf{n}, 
\end{equation}
where measurement vector $\mathbf{y} \in \mathbb{C}^{T \times 1}$, sensing matrix $\mathbf{S} \in \mathbb{C}^{T \times N}$, and $\mathbf{n}\in {{\mathbb{C}}^{T\times 1}}\sim\mathcal{C}\mathcal{N}(0,{{\beta }^{-1}}{{\mathbf{I} }_{T}})$ denotes an AWGN vector. It assumes that the elements in $\mathbf{x}$ are independent, and the following two-layer sparsity-promoting prior is employed
\begin{equation}
p\left( \mathbf{x} \middle| \boldsymbol{\gamma} \right) = {\prod\limits_{n}{p\left( x_{n} \middle| \gamma_{n} \right)}} = {\prod\limits_{n}{\mathcal{N}\left( x_{n} \middle| 0,{\gamma_{n}}^{- 1} \right)}},
\end{equation}
\begin{equation}
p\left( \boldsymbol{\gamma} \right) = {\prod\limits_{n}{p\left( \gamma_{n} \right)}} = {\prod\limits_{n}{Ga\left( \gamma_{n} \middle| \epsilon,\eta \right)}},
\end{equation}
where $\epsilon$ and $\eta$ are shape parameter and rate parameter of the Gamma distribution, respectively. In the SBL algorithm, the precision vector $\boldsymbol{\gamma}$
%Sparse Bayesian learning (SBL) is another promising algorithm for solving these kinds of CS issues, which shows advantages in terms of performance. 
%many approaches have been proposed to deal with it, including orthogonal matching pursuit (OMP) \cite{RN17}, iterative reweighted method \cite{RN18} and manifold optimisation (MO) \cite{RN19}. 
%In the basic SBL algorithm \cite{RN20}, the precision vector $
%\boldsymbol{\gamma} \in \mathbb{C}^{N \times 1}$ is introduced, each of its elements corresponds to an element of the sparse recovery vector $\mathbf{x}$. When an element ${\gamma}_{n}$ approaches zero, its corresponding ${x}_{n}$ can be considered as a non-zero point. Assuming each element in $\mathbf{x}$ is independent, the two-layer sparsity-promoting prior is employed:
learned by maximizing the posterior distribution
\begin{equation}
p\left( \boldsymbol{\gamma} \middle| \mathbf{y} \right) \propto p\left( \mathbf{y} \middle| \boldsymbol{\gamma} \right)p\left( \boldsymbol{\gamma} \right). 
\end{equation}
As the value of $\boldsymbol{\gamma}$ cannot be obtained in a closed form, an iterative re-estimation method is used, where the following covariance matrix $\mathbf{\Sigma}$  and mean vector $\boldsymbol{\mu}$  need to be computed in each iteration 
\begin{equation}
\mathbf{\Sigma} = \left\lbrack \beta\mathbf{S}^{H}\mathbf{S} + \mathbf{d}\mathbf{i}\mathbf{a}\mathbf{g}\left( \boldsymbol{\gamma} \right) \right\rbrack^{- 1},
\end{equation}
\begin{equation}
\boldsymbol{\mu} = \beta\mathbf{\Sigma}\mathbf{S}^{H}\mathbf{y},
\end{equation}
where  $\beta$ is the noise precision. In each iteration, the precision $\gamma_n$ is updated as
\begin{equation}
\gamma_n= (2\epsilon+1)/(2\eta+ {| \mu_{n} |}^2+\Sigma_{n,n}), n=1,2,...,n
\end{equation}
and the estimate of $\mathbf{x}$ is given by $\boldsymbol{\mu}$. Due to the matrix inverse operation in (18) in each iteration, the complexity of the SBL algorithm is high.    

To overcome the problem, the UAMP-SBL algorithm was proposed in \cite{RN16}, leveraging variational inference \cite{RN90,RN91,RN92} and UAMP \cite{RN21}. The sparse signal recovery problem is represented by a factor graph model, based on which approximate inference is implemented with variational message passing, and UAMP is incorporated to the message passing to handle the most computationally intensive part of message computations with low complexity and high robustness. To facilitate the incorporation of UAMP, a unitary transformation is performed to the original model (14). Specifically, with the SVD of the sensing matrix
\begin{equation}
\textbf{S}=\mathbf{U}\Breve{\mathbf{S}}{\mathbf{V}}^{H},    
\end{equation}
the transformed model can be represented as
\begin{equation}
\mathbf{z} = \boldsymbol{\psi}\mathbf{x} + \Breve{\mathbf{n}}
\end{equation}
where $
\mathbf{z}={\mathbf{U}}^{H}\mathbf{y}$, $\boldsymbol{\psi}=\Breve{\mathbf{S}}{\mathbf{V}}^{H}={\mathbf{U}}^{H}\mathbf{S}$, and $\Breve{\mathbf{n}}={\mathbf{U}}^{H}\mathbf{n}$. 

\begin{algorithm}[H]
\caption{UAMP-SBL}\label{alg:alg1}
\textbf{Model:} $
\mathbf{z} = {\mathbf{U}}^{H}\mathbf{y} = {\mathbf{U}}^{H}\mathbf{U}\Breve{\mathbf{S}}{\mathbf{V}}^{H}\mathbf{x} + \Breve{\mathbf{n}} = \boldsymbol{\psi}\mathbf{x} + \Breve{\mathbf{n}}$

\textbf{Initialization:} $\boldsymbol{\varpi} = \Breve{\mathbf{S}}{\Breve{\mathbf{S}}}^{H}\mathbf{1}$, threshold $\delta_{th}=0.0001$, ${t}_{x}^{(0)} = 1,~\mathbf{x}^{(0)} = \mathbf{0},~\mathbf{e}^{(-1)} = \mathbf{0},$ ${\beta}^{(-1)}=1,$ ${\epsilon}^{(0)}=0.01,$ $\boldsymbol{\gamma}^{(0)}=\mathbf{1},$ $i=0$ and $I$ is the preset total number of iterations

\begin{algorithmic} [1]
\WHILE{$i\leq I$ and
     $\left\| {\mathbf{x}^{i + 1} - \mathbf{x}^{i}} \right\|^{2}/\left\| \mathbf{x}^{i + 1} \right\|^{2} > \delta_{th}$}

  \STATE $
  \mathbf{t}_{p} = t_{x}^{i}\boldsymbol{\varpi}$ 

  \STATE $ 
  \mathbf{p} = \boldsymbol{\psi}\mathbf{x}^{i} - \mathbf{t}_{p} \cdot \mathbf{e}^{i - 1} $

  \STATE $ 
  \mathbf{v}_{r} = \mathbf{t}_{p}./\left( \mathbf{1} + \beta^{i - 1}\mathbf{t}_{p} \right) $

  \STATE $ 
  \mathbf{r} = \left( {\beta^{i - 1}\mathbf{t}_{p} \cdot \mathbf{z} + \mathbf{p}} \right)./\left( \mathbf{1} + \beta^{i - 1}\mathbf{t}_{p} \right) $

  \STATE $ 
  \beta^{i} = T/\left( \left\| \mathbf{z} - \mathbf{r} \right\|^{2} + \mathbf{1}^{H}\mathbf{v}_{r} \right) $

 \STATE $
  \mathbf{t}_{s} = \mathbf{1}./\left( \mathbf{t}_{p} + \left( \beta^{i} \right)^{-1}\mathbf{1} \right)$

  \STATE $
  \mathbf{e}^{i} = \mathbf{t}_{s} \cdot \left( {\mathbf{z} - \mathbf{p}} \right)$

  \STATE $
  \left. t_{q} = N/\left( \boldsymbol{\varpi} \right.^{H}\mathbf{t}_{s} \right)$

  \STATE $
  \mathbf{q} = \mathbf{x}^{i} + t_{q} \cdot \left( \boldsymbol{\psi}^{H}\mathbf{e}^{i} \right)$

  \STATE $
  t_{x}^{i+1} = \left( t_{q}/N \right) \cdot
  \mathbf{1}^{H}\left\lbrack \mathbf{1}./\left( \mathbf{1} + t_{q}\boldsymbol{\gamma}^{i} \right) \right\rbrack$

  \STATE $
  \mathbf{x}^{i+1} = \mathbf{q}./\left( \mathbf{1} + t_{q}\boldsymbol{\gamma}^{i} \right)$

  \STATE $
  \forall n:
        \gamma_{n}^{(i + 1)} = \left( 2\epsilon^{i} + 1 \right)/\left( \| x_{n}^{(i + 1)}\|^{2} + t_{x}^{(i + 1)} \right)$
        
  \STATE $
  \small{\epsilon^{i + 1} = \frac{1}{2}\sqrt{{\log\left\lbrack {\frac{1}{N}\left( {\sum\limits_{n}\gamma_{n}^{i + 1}} \right)} \right\rbrack} - \frac{1}{N}\left\lbrack {\sum\limits_{n}{log\left( \gamma_{n}^{i + 1} \right)}} \right\rbrack}}$
  
  \STATE $
  i=i+1$
\ENDWHILE

\end{algorithmic}
\label{alg1}
\end{algorithm}

Then an auxiliary variable $\mathbf{r} = {\boldsymbol{\psi}\mathbf{x}}$ is introduced, which is crucial to an efficient realization of SBL. The conditional joint distribution of variables can be represented as
%\begin{equation}
%p\left( \mathbf{x},\boldsymbol{\gamma},\beta|\mathbf{z} \right) \propto %p\left( \mathbf{z}|\mathbf{x},\beta \right)p\left( %\mathbf{x}|\boldsymbol{\gamma} \right)p(\boldsymbol{\gamma})p(\beta),
%\end{equation}
\begin{equation}
\begin{aligned}
p( \mathbf{x},\mathbf{r}, & \boldsymbol{\gamma},\beta|\mathbf{z} )  \propto p\left( \mathbf{z}|\mathbf{r},\beta \right)p\left( \mathbf{r}|\mathbf{x} \right)p\left( \mathbf{x}|\boldsymbol{\gamma} \right)p\left( \boldsymbol{\gamma}|\epsilon \right)p(\beta) \\ 
& =
\prod\limits_{m = 1}^{M}{\left. \mathcal{N}\left( z_{m} \right|r_{m},\beta^{- 1} \right){\prod\limits_{m = 1}^{M}{\delta\left( r_{m} - \left\lbrack \boldsymbol{\psi} \right\rbrack_{m}\mathbf{x} \right)}}} \\
& \ \ \ \ \prod\limits_{n = 1}^{N}{\left. \mathcal{N}\left( x_{n} \right|0,{\gamma_{n}}^{- 1} \right){\prod\limits_{n = 1}^{N}{Ga\left( \gamma_{n} \middle| \epsilon,\eta \right)p(\beta)}}}
\end{aligned}
\end{equation}

With the structured trial distribution
\begin{equation}
q(\mathbf{x},\mathbf{r},\boldsymbol{\gamma},\beta)=q(\beta)q(\mathbf{x},\mathbf{r})q(\boldsymbol{\gamma}),
\end{equation}
variational inference can be performed and implemented with message passing.
Moreover, the impact of the shape parameter $\epsilon$ on the sparse signal recovery is analyzed in \cite{RN16}, and a rule for automatically tuning the shape parameter is found, i.e., 
\begin{equation}
\epsilon = \frac{1}{2}\sqrt{{\log\left\lbrack {\frac{1}{N}\left( {\sum_{n}\gamma_{n}} \right)} \right\rbrack} - \frac{1}{N}\left\lbrack {\sum_{n}{log\left( \gamma_{n} \right)}} \right\rbrack}.
\end{equation}

It is also noted that, as the noise precision is normally unknown, its estimation is also incorporated into the algorithm. The UAMP-SBL algorithm is summarized in Algorithm 1.

\subsection{Proposed Channel Estimator for Scenario 1 }
Regarding the CS-based channel estimation problem for the RIS-assisted communication system formulated in (13), the plain UAMP-SBL algorithm designed for a typical SMV problem does not have the mechanism to exploit the structure of the matrices, which however is crucial to improving the channel estimation performance. This motivates us to design the dedicated UAMPSBL-PCI algorithm.
%However, considering that (13) is a MMV model, and the addressed cascaded channel has a special dual-structure sparsity, simply applying the plain UAMP-SBL algorithm to this model not only greatly increases the computational cost, but also leads to a degradation in estimation performance. Therefore, we propose a new channel estimation scheme, named UAMP-SBL with partial common support identification (UAMPSBL-PCI).

Compared to the plain UAMP-SBL algorithm, UAMPSBL-PCI is designed for the formulated MMV problem to exploit the dual-structure sparsity of the channel effectively. Specifically, as in \cite{RN10}, the sparse common row support is firstly acquired, leading to significant reduction in the size of the CS problem. Furthermore, UAMPSBL-PCI makes use of a fast scan mechanism to obtain the partially common column support, and the knowledge is then exploited in the sparse channel matrix estimation. %In addition, the UAMPSBL-PCI algorithm is modified to be suitable for the MMV model. 
In the following, we elaborate the proposed estimation algorithm.  

\begin{algorithm}[H]
\caption{Common row support acquisition}\label{alg:alg2}
   \textbf{Define:}\ Function $Ind\left( a_{i} \right)$ returns the index information of elements $a_{i}$ in $\mathbf{a}$.  \\
   \textbf{Input:} $\Breve{\mathbf{Y}}_{j},J,M,{P}_{BR}$
    \\
    \textbf{Initialization:} $\mathbf{r} = \mathbf{0}$
    \\
   \textbf{Output:}\ $\mathbb{R}$
    \begin{algorithmic} [1]
       \FOR{$m=1:M$}
        \STATE $
        \mathbf{r}(m) =  \sum\limits_{j = 1}^{J}\left\| {\Breve{\mathbf{Y}}_{j}(:,m)} \right\|^{2}$ 
        \ENDFOR
    \STATE $\mathbb{R}=Ind\left( \mathop{max}\limits_{P_{BR}}\mathbf{r}  \right)$
\end{algorithmic}
\label{alg2}
\end{algorithm}

First, the common row support acquisition is presented in Algorithm 2. The previous analysis indicates that the cascaded channels of each user $[\Breve{\mathbf{H}}_{j}]_{j = 1}^{J}$ have a completely common row sparsity in the RIS-assisted communication system. From the perspective of the channel matrices, all non-zero elements are located in the same row. Let $\mathbb{R}_{j}$ denote the row set of $\Breve{\mathbf{H}}_{j}$ which contains all non-zero elements, then we have
\begin{equation}
\mathbb{R}_{1}=\mathbb{R}_{2}=...=\mathbb{R}_{j}.
\end{equation}

It should be noted that a conjugate transpose is performed on $\Breve{\mathbf{H}}_{j}$ to formulate the CS-based model (13). Thus, the row sparsity of the channel matrix manifests itself in the columns of the measurement matrix $\Breve{\mathbf{Y}}_{j}$. Due to the fact that only non-zero elements are capable of contributing power to the received signal, we introduce an indicator
\begin{equation}
 \mathbf{r}(m) =  \sum\limits_{j = 1}^{J}\left\| {\Breve{\mathbf{Y}}_{j}(:,m)} \right\|^{2} 
\end{equation}
which is the sum power of column $m$ of the measurement matrix in steps 1-3.

Suppose that there exist ${P}_{BR}$ transmission paths between the BS and RIS, the columns corresponding to the ${P}_{BR}$ elements with the largest magnitude of indicator vector $\mathbf{r}$ are considered as the common row support $\mathbb{R}$, which is acquired by
\begin{equation}
\mathbb{R}=Ind\left( \mathop{max}\limits_{P_{BR}}\mathbf{r}  \right)
\end{equation}
as shown in step 4. Hence, only the rows indicated by $\mathbb{R}$ in the cascaded channel $[\Breve{\mathbf{H}}_{j}]_{j = 1}^{J}$ need to be estimated, which greatly reduces the computational cost of channel estimation algorithms.

Then, based on the common row support $\mathbb{R}$ obtained above, we can design the UAMPSBL-PCI algorithm as shown in Algorithm 3. Initially, for each common row $\alpha$, we perform a unitary transformation to the corresponding measurement matrix:
 \begin{equation}
\forall j: \mathbf{Z}_{\alpha}(:,j)=\mathbf{U}^{H}{\Breve{\mathbf{Y}}_{j}}(:,\mathbb{R}(\alpha)).
\end{equation}
Following the plain UAMP-SBL algorithm, we update the variable matricies
$\mathbf{T_p}$, $\mathbf{P}$, $\mathbf{V_r}$ and $\mathbf{R}$ accordingly as shown in steps 4-7. In step 8, the noise precision ${\boldsymbol{\beta}(j)}$ for each user in the $i$-th iteration can be updated as:
 \begin{equation}
\forall j:
        \small{{\boldsymbol{\beta}(j)}^{(i)} = T/\left\lbrack \left\| \mathbf{Z}_{\alpha}(:,j) - \mathbf{R}(:,j) \right\|^{2} + \mathbf{1}^{H}\mathbf{V_r}(:,j)\right\rbrack}.
\end{equation}
Afterwards, the mean vector $\mathbf{t_q}$ and variance matrix $\mathbf{Q}$ is calculated in steps 9-12 while the temporary estimated matrix $\mathbf{X}$ used in the next iteration is obtained in steps 13-14. Furthermore, each element of the precision matrix $\boldsymbol{\gamma}$ in the $(i+1)$-th iteration can be obtained as 

\begin{equation}
\begin{aligned}
\forall j, \forall n:
        \boldsymbol{\gamma}(n,j)^{(i+1)}  = & \left\lbrack 2{\boldsymbol{\epsilon}(j)}^{(i)} + 1 \right\rbrack / \\ &\left\lbrack \| {\mathbf{X}(n,j)}^{(i+1)}\|^{2} + {\mathbf{{t}_{x}}(j)}^{(i+1)} \right\rbrack.    
\end{aligned}
\end{equation}

It is known that the elements in the precision matrix $\boldsymbol{\gamma}$ bear an one-to-one correspondence with the recovery matrix $\mathbf{X}$. A large value of the precision matrix element ${\gamma}_{n}$ indicates that the corresponding recovery matrix element ${x}_{n}$ is zero. Thus, in order to exploit the partially common column sparsity of the channel matrices, we introduce a fast scan mechanism in UAMPSBL-PCI as shown in steps 16-21. Specifically, with the estimation results of the initial $I_{fs}$ iterations, it can roughly delineate the positions where the non-zero elements exist. By incorporating these information into the subsequent iterations, a substantial enhancement in estimation performance can be achieved.

Suppose there are ${P}_{j}$ transmission paths between each user and RIS, and only ${P}_{j}$ non-zero elements exist on each common row for each user. Thus, the index corresponding to the smallest ${P}_{j}$ elements in each column of $\boldsymbol{\gamma}$, which indicates the columns in $[\Breve{\mathbf{H}}_{j}(\alpha,:)]_{j = 1}^{J}$ with non-zero elements, are denoted as 
\begin{equation}
\forall j:
 \mathbf{D}(:,j) =  Ind\left(\mathop{min}\limits_{P_j}\boldsymbol{\gamma}(:,j)^{(i+1)}\right).
\end{equation}
In Scenario 1, it is assumed that all users share some common paths. Thus, it is expected that the sparse columns of each user exhibit overlap. These overlapping columns can be considered as the partial common column support. Let $\mathbb{C}^{j}_{\alpha}$ denotes the column set of the $\alpha$-th common row of $\Breve{\mathbf{H}}_{j}$ that contains non-zero element, we have
\begin{equation} 
\mathbb{C}^{1}_{\alpha}\cap\mathbb{C}^{2}_{\alpha}\cap...\cap\mathbb{C}^{j}_{\alpha}=\mathbb{C}^{Com}_{\alpha}, \alpha=1,2,..,{P}_{BR}.
\end{equation} 
Their identification is determined through counting the frequency of  elements appears in variable $\mathbf{D}$. Specifically, if there are $P_c$ common columns among the users,
\begin{equation} 
\mathbb{C}_{\alpha}^{Com} = \mathop{maxnum}\limits_{P_c}\mathbf{D}.
\end{equation} 
In the subsequent iterations, the precision matrix $\boldsymbol{\gamma}$ is updated as
\begin{equation} 
\forall n\in\mathbb{C}_{\alpha}^{Com}:
\boldsymbol{\gamma}(n,:)^{(i+1)} = \frac{1}{J}\sum\limits_{j=1}^{J}\boldsymbol{\gamma}(n,j)^{(i+1)}
\end{equation} 
according to the obtained sparse prior.

Finally, we can update the shape parameter $\boldsymbol{\epsilon}$ following the plain UAMP-SBL algorithm in each iteration as in step 22, and acquire the final estimation result ${\Breve{\mathbf{H}}_{j}}^{H}$ in step 24.

\subsection{Proposed Channel Estimator for Scenario 2 }

In this subsection, we extend the proposed UAMPSBL-PCI algorithm to handle the more general Scenario 2 where subsets of users share the common channels.

Various techniques have been studied to tackle the challenging problem of RIS channel estimation. However, these techniques often simplify the system model by assuming a simple scenario where all users share partial common channels, as described in Scenario 1. In practical settings, it is uncertain whether there exists a shared transmission path between users, as depicted in Scenario 2, which leads to difficulty in channel estimation. To address this issue, we propose a novel approach that utilizes the UAMPSBL-PCI algorithm with an auto-clustering mechanism. It can be realized by replacing the steps 16-21 with Algorithm 4.

\begin{algorithm}[H]
\caption{UAMPSBL-PCI}\label{alg:alg3}
   \textbf{Input:} $\Breve{\mathbf{Y}}_{j},\Breve{\boldsymbol{\Omega}},J,{P}_{BR},{P}_{j},{P}_{c},N,T,\mathbb{R}$
    \\
    \textbf{Initialization:}\ threshold $\delta_{th}=0.0001$, $i=0$, $I_{fs}$ is the iterations for fast scan and $I$ is the preset total number of iterations,
    $\Breve{\boldsymbol{\Omega}}=\mathbf{U}\boldsymbol{\omega}{\mathbf{V}}^{H}$, $\boldsymbol{\varpi}= \boldsymbol{\omega}\boldsymbol{\omega}^{H}\mathbf{1}$, $\boldsymbol{\psi}={\mathbf{U}}^{H}{\Breve{\mathbf{\Omega}}}$, $\forall j:{\Breve{\mathbf{H}}}_{j}^{H}=\mathbf{0}$,$~{\mathbf{{t}_{x}}(j)}^{(0)} = 1$,$~{\mathbf{X}(:,j)}^{(0)} = \mathbf{0}$, ${\mathbf{E}(:,j)}^{(-1)} = \mathbf{0}$, ${\boldsymbol{\beta}(j)}^{(-1)}=1$, ${\boldsymbol{\epsilon}(j)}^{(0)}=1$, ${\boldsymbol{\gamma}(:,j)}^{(0)}=\mathbf{1}$, $\mathbb{C}^{Com}=\mathbf{0}$
    
   \textbf{Output:}\ ${\Breve{\mathbf{H}}_{j}}^{H}$
   
\begin{algorithmic} [1]

  \FOR{$\alpha=1:{P}_{BR}$}
    
      \STATE $\forall j: \mathbf{Z}_{\alpha}(:,j)=\mathbf{U}^{H}{\Breve{\mathbf{Y}}_{j}}(:,\mathbb{R}(\alpha))$
      \WHILE{ $\sum\limits_{j = 1}^{J}\left\| {{\mathbf{X}(:,j)}^{(i + 1)} - {\mathbf{X}(:,j)}^{(i)}} \right\|^{2}/\sum\limits_{j = 1}^{J}\left\| {\mathbf{X}(:,j)}^{(i + 1)} \right\|^{2} $ \\ $> \delta_{th}$ and $i\leq I$
      }

        \STATE $
        \small{\mathbf{T_p} = \boldsymbol{\varpi}{\mathbf{{t}_{x}}}^{(i)}}$ 

        \STATE $
        \small{\mathbf{P} = \boldsymbol{\psi}{\mathbf{X}}^{(i)} - \mathbf{T_p} \cdot {\mathbf{E}}^{(i-1)}} $

        \STATE $ 
        \small{\mathbf{V_r} = \mathbf{T_p}./\left\lbrack \mathbf{1} + \mathbf{T_p}~\text{diag}({\boldsymbol{\beta}}^{(i-1)})\right\rbrack} $

        \STATE $
        \small{\mathbf{R} = \left\lbrack {\mathbf{T_p}~\text{diag}({\boldsymbol{\beta}}^{(i-1)}) \cdot \mathbf{Z}_{\alpha} + \mathbf{P}}\right\rbrack  ./ \left\lbrack \mathbf{1} + \mathbf{T_p}~\text{diag}({\boldsymbol{\beta}}^{(i-1)}) \right\rbrack } $

        \STATE $ \forall j:
        \small{{\boldsymbol{\beta}(j)}^{(i)} = T/\left\lbrack \left\| \mathbf{Z}_{\alpha}(:,j) - \mathbf{R}(:,j) \right\|^{2} + \mathbf{1}^{H}\mathbf{V_r}(:,j)\right\rbrack}$

        \STATE $
        \small{\mathbf{T_s} = \mathbf{1}./\left\lbrack \mathbf{T_p} + \mathbf{1} \left( \mathbf{1}./{\boldsymbol{\beta}}^{(i)} \right) \right\rbrack}$

        \STATE $
        \small{\mathbf{E}^{(i)} = \mathbf{T_s} \cdot \left( {\mathbf{Z}_{\alpha} - \mathbf{P}} \right)}$

        \STATE $
        \mathbf{t_q} = N./\left( \boldsymbol{\varpi}^{H}\mathbf{T_s} \right)$

        \STATE $
        \small{\mathbf{Q} = {\mathbf{X}}^{(i)} + (\boldsymbol{\psi}^{H}\mathbf{E}^{(i)})~\text{diag}(\mathbf{t_q})   }$

        \STATE $
        \small{{\mathbf{{t}_{x}}}^{(i+1)} = \left( \mathbf{t_q}/N \right) \cdot
        \mathbf{1}^{H}\left\lbrack \mathbf{1}./\left( \mathbf{1} + \boldsymbol{\gamma}^{(i)}~\text{diag}(\mathbf{t_q}) \right) \right\rbrack}$

        \STATE $
        \small{{\mathbf{X}}^{(i+1)} = \mathbf{Q}./\left( \mathbf{1} + \boldsymbol{\gamma}^{(i)}~\text{diag}(\mathbf{t_q}) \right)}$

        \STATE $\forall j, \forall n:
        \small{\boldsymbol{\gamma}(n,j)^{(i+1)} = \left\lbrack 2{\boldsymbol{\epsilon}(j)}^{(i)} + 1 \right\rbrack/}$ \\ \ \ \ \ \ \ \ \ \ \ \ \ \ \ \ \ \ \ \ \ \ \ \ \ \ \ \ \ \ \ $\small{\left\lbrack \| {\mathbf{X}(n,j)}^{(i+1)}\|^{2} + {\mathbf{{t}_{x}}(j)}^{(i+1)} \right\rbrack}$

        \IF{$i=I_{fs}$}
                \STATE $\forall j:
                \mathbf{D}(:,j) =  Ind\left(\mathop{min}\limits_{P_j}\boldsymbol{\gamma}(:,j)^{(i+1)}\right)  $

                \STATE $\mathbb{C}_{\alpha}^{Com} = \mathop{maxnum}\limits_{P_c}\mathbf{D}$
        \ELSIF{$i>I_{fs}$}
                \STATE $\forall n\in\mathbb{C}_{\alpha}^{Com}:
                \boldsymbol{\gamma}(n,:)^{(i+1)} = \frac{1}{J}\sum\limits_{j=1}^{J}\boldsymbol{\gamma}(n,j)^{(i+1)}$
        \ENDIF

       \STATE $\forall j:
       \small{{\boldsymbol{\epsilon}(j)}^{(i+1)} = \frac{1}{2}\sqrt{{\log\left\lbrack \frac{ {\sum\limits_{n}\boldsymbol{\gamma}(n,j)^{(i+1)}} }{N} \right\rbrack} - \frac{ {\sum\limits_{n}{log\left\lbrack \boldsymbol{\gamma}(n,j)^{(i+1)} \right\rbrack}} }{N}}}$
    \ENDWHILE
    \STATE $\forall j: {\Breve{\mathbf{H}}}_{j}^{H}(:,\mathbb{R}(\alpha))=\mathbf{X}(:,j)^{(i+1)}$
   
  \ENDFOR 
\end{algorithmic}
\label{alg3}
\end{algorithm}

In this more general scenario, it is still reasonable to assume that all users share the same row support, thereby Algorithm 2 can still be used to acquire the common row support. Fig. 4 illustrates that while there is no common column shared by all users on any common row, there exist some common columns among a subset of users, which are critical for optimizing the estimation algorithm by leveraging sparsity. %These special columns are crucial for optimizing the estimation algorithm by utilizing the sparsity feature. 
To identify the positions of these special columns, an auto-clustering mechanism is developed.

When the element ${\gamma}_{n}$ in the precision matrix is large, the corresponding estimated element ${x}_{n}$ can be regarded as a zero element. In the first step, the minimum value in $\boldsymbol{\gamma}$ is used as the threshold ${\delta}_{\alpha}$:
\begin{equation} 
{\delta}_{\alpha}=min(\boldsymbol{\gamma}^{(i+1)}).
\end{equation}

Next, each row of the precision matrix is sorted in ascending order. If the minimum value of one row is significantly larger than the threshold ${\delta}_{\alpha}$, it implies there is no non-zero element in the corresponding columns of estimated cascaded channel $[\Breve{\mathbf{H}}_{j}]_{j = 1}^{J}$ as shown in steps 4-7. Otherwise, the elements in the row are compared sequentially. When the next element is much greater than the average value of the previous sequence, the comparison stops, and all the elements thus far are considered non-zero as shown in steps 8-11. In step 12, we then cluster these elements and assign them an average value, obtaining the partial common support $\mathbb{C}_{\alpha}$ for each common row by reorganizing the comparison results as
\begin{equation} 
\mathbb{C}_{\alpha}(n,\mathbf{R}(1:j))={b}_{t}/j.
\end{equation}

Finally, incorporating the partial common support $\mathbb{C}_{\alpha}$ into the subsequent iterations:
\begin{equation} 
\forall \mathbb{C}_{\alpha}(n,j)\neq 0:
\boldsymbol{\gamma}(n,j)^{(i+1)}=\mathbb{C}_{\alpha}(n,j).
\end{equation} 
These lead to the UAMPSBL-PCI channel estimator with auto-clustering mechanism.

\begin{algorithm}[H]
\renewcommand{\thealgorithm}{4}
\caption{Auto-clustering mechanism}\label{alg:alg4}
   \textbf{Define:}\ Operator $F[\mathbf{a}]$ sorts vector $\mathbf{a}$ in ascending order. $V_{1}$ and $V_{2}$ are preset magnification coefficients.\\
   \textbf{Initialization:} $\mathbb{C}_{\alpha}=\mathbf{0}$
    \begin{algorithmic} [1]
        \IF{$i=I_{fs}$} 
        \STATE ${\delta}_{\alpha}=min(\boldsymbol{\gamma}^{(i+1)})$
        \FOR{$n=1:N$}
            \STATE $j=1; \mathbf{q}= F\left\lbrack\boldsymbol{\gamma}(n,:)^{{i+1}}\right\rbrack; \mathbf{R}=Ind(\mathbf{q})$
            \IF{$\mathbf{q}(1)>V_{1}\cdot{\delta}_{t}$}
                \STATE \textbf{Continue}
            \ENDIF
            \STATE ${b}_{t}={b}_{c}=\mathbf{q}(1)$
            \WHILE{$j\leq(J-1)$ and $\mathbf{q}(j+1)<V_{2}\cdot{b}_{c}$}
                
                \STATE ${b}_{t}={b}_{t}+\mathbf{q}(j+1); j=j+1;{b}_{c}={b}_{t}/j$
            
            \ENDWHILE
        \STATE $\mathbb{C}_{\alpha}(n,\mathbf{R}(1:j))={b}_{t}/j$
        \ENDFOR
        \ELSIF{$i>I_{fs}$}
                \STATE $\forall \mathbb{C}_{\alpha}(n,j)\neq 0:
                \boldsymbol{\gamma}(n,j)^{(i+1)}=\mathbb{C}_{\alpha}(n,j)
                $
        \ENDIF
        
\end{algorithmic}
\label{alg4}
\end{algorithm}

\subsection{Complexity analysis}

In this subsection, we analyze the complexity of the proposed UAMPSBL-PCI algorithm. It can be seen that UAMPSBL-PCI consists of two parts, and their complexity will be analyzed separately. For the common row support acquisition as shown in Algorithm 2, we compute the sum power of $J$ measurement matrices of dimension $T\times M$. Therefore, its complexity can be calculated as $\mathcal{O}(JTM)$. The main part of the UAMPSBL-PCI as shown in Algorithm 3 is modified from the plain UAMP-SBL algorithm. Considering the dimension $T\times N$ of the measurement matrix, the complexity of one UAMP-SBL iteration is $\mathcal{O}(TN)$ \cite{RN16}. In Algorithm 3, we perform the estimation on ${P}_{BR}$ common columns of $J$ users respectively. Given the preset number of iterations for each estimation is $I$, the total computational complexity is $\mathcal{O}(JTN{P}_{BR}I)$. Furthermore, the UAMPSBL-PCI algorithm incorporates two distinct fast scan mechanism (Steps 16-21 of Algorithm 3 and Algorithm 4) according to different scenarios. However, since there is no multiplication and division operation involved,  making their complexity negligible. Therefore, the total complexity of the proposed UAMPSBL-PCI algorithm is $\mathcal{O}(JTM)+\mathcal{O}(JTN{P}_{BR}I)$. 
The analysis is summarized in Table 1.

\section{Simulation results}
In this section, we conduct extensive simulations to evaluate the channel estimation performance of the proposed UAMPSBL-PCI algorithm for the RIS-assisted communication system. The proposed algorithm is compared to a number of different channel estimation techniques, such as conventional OMP \cite{RN22}, DS-OMP \cite{RN10},  S-MJCE \cite{RN23} and UAMP-SBL \cite{RN16}. The performance of the Oracle Least Squares (LS) based technique serves as a  lower bound, where the sparse position of the cascaded channel is assumed to be known. The conventional OMP algorithm utilizes classical CS model processing methods to estimate the sparse cascaded channel matrix. Both DS-OMP and S-MJCE algorithms exploit the dual-structure sparsity of the cascaded channel in a RIS-assisted system. The former performs joint OMP estimation of common rows and partial common columns, and then separately estimates non-common columns for each user in turn to obtain the required CSI. The latter first performs subspace estimation based on column sparsity, projects the received signal into an estimated subspace, and then utilizes row sparsity in iterative reweighted-based alternative optimization for multiple users to estimate the channel matrix.

%\begin{table}[!t]
\begin{table}
\caption{Complexity analysis for UAMPSBL-PCI\label{tab:table1}}
\centering
\begin{tabular}{|c||c|}
\hline
Row support acquisition & $\mathcal{O}(JTM)$\\
\hline
Main part of UAMPSBL-PCI & $\mathcal{O}(JTN{P}_{BR}I)$\\
\hline
fast scan mechanism & $0$\\
\hline
Total & $\mathcal{O}(JTM)+\mathcal{O}(JTN{P}_{BR}I)$\\
\hline
\end{tabular}
\end{table}

We assume the number of users $J=16$, the number of UPA antennas $M=8\times8$ at the BS, and the number of units $N=16\times16$ at the RIS. The default SNR is $0$dB, while the pilot overhead $T=192$. It is assumed that there are five paths between the BS and RIS, and each user has ten paths toward the RIS $(P_{BR}=5,P_{j}=10)$. The channel gain $\alpha_{\rho}^{BR}$ and $\alpha_{\varsigma}^{j}$ are set as $10^{-3}d_{BR}^{-2.2}$ and $10^{-3}d_{j}^{-2.8}$ respectively. The distances between BS-RIS and RIS-User are set to be $10m$ and $100m$ respectively \cite{RN24}. The AoA and AoD parameters of each RIS element are uniformly distributed over $[-\pi/2, \pi/2)$ while lying on the discretized grid. The normalized mean square error (NMSE), which is defined as $
\mathbb{E}\left\lbrack \left\| {{{\Breve{\mathbf{H}}}_{j}}^{H} - {\mathbf{H}_{j}}^{H}} \right\|_{F}^{2}/\left\| {\mathbf{H}_{j}}^{H} \right\|_{F}^{2} \right\rbrack$, is used to evaluate the channel estiamtion performance. Monte Carlo experiments are performed to reduce random errors.

\subsection{Simulation results for Scenario 1 }

We first present channel estimation performance of the proposed UAMPSBL-PCI algorithm for the RIS-assisted communication system in Scenario 1 where all users share the common channels.

\begin{figure}[!t]
\centering
\includegraphics[width=3.5in]{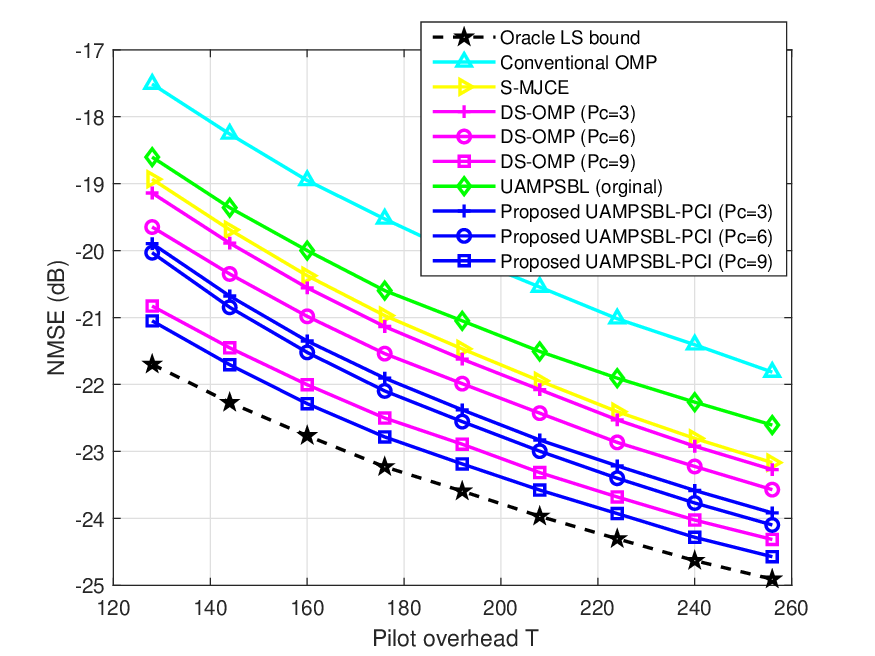}
\caption{Estimation performance NMSE against the pilot overhead $T$ in Scenario 1.}
\label{fig_5}
\end{figure}

\begin{figure}[!t]
\centering
\includegraphics[width=3.5in]{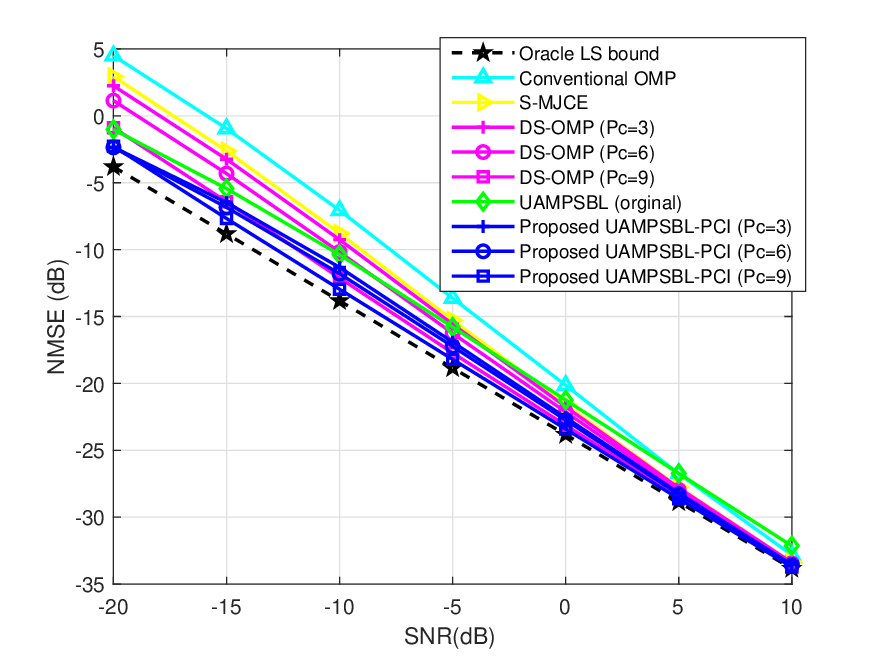}
\caption{Estimation performance NMSE against the SNR in Scenario 1.}
\label{fig_6}
\end{figure}

Fig. 5 shows channel estimation performance against pilot overhead $T$. There are several observations regarding the simulation results. First, it is evident that the accuracy of all channel estimation algorithms improves with increasing pilot overhead. Second, as the number of partial common columns $P_{c}$ increases, both the DS-OMP and proposed UAMPSBL-PCI algorithms show improved estimation performance, mainly due to more precise access to the sparse position. Moreover, to accomplish the same estimation accuracy under the same system conditions, the proposed UAMPSBL-PCI algorithm requires the lowest amount of pilot overhead, proving its superiority.

Fig. 6 shows the channel estimation performance against SNR. As the SNR increases from -20 dB to 10 dB, it is easy to observe that the estimation accuracy of all channel estimation algorithms improves and approaches the theoretical limit as the channel conditions continue to improve. Among them, the proposed UAMPSBL-PCI algorithm exhibits great performance advantages and robustness, especially in poor channel conditions. This implies that it is more appropriate for challenging, constantly changing situations in reality.

\begin{figure}[!t]
\centering
\includegraphics[width=3.5in]{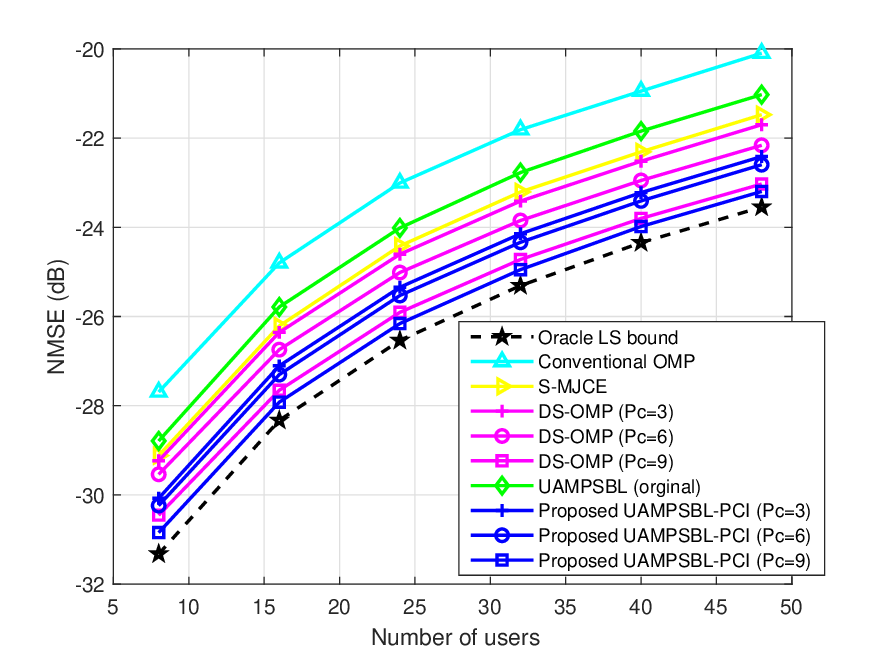}
\caption{Estimation performance NMSE against the number of users $J$ in Scenario 1.}
\label{fig_7}
\end{figure}

\begin{figure}[!t]
\centering
\includegraphics[width=3.5in]{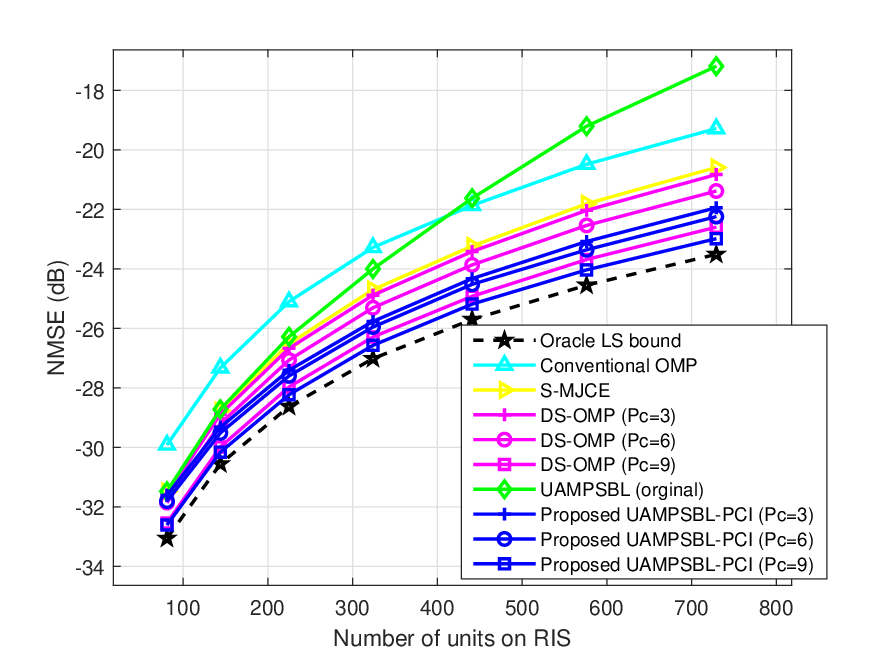}
\caption{Estimation performance NMSE against the number of units $N$ on RIS in Scenario 1. }
\label{fig_8}
\end{figure}

Fig. 7 shows the channel estimation performance against the number of users $J$. As the number of users grows, the estimation performance of all algorithms decreases, mainly due to the increase in non-common column estimation elements and the overall system complexity. However, it is noteworthy that under identical system settings, the proposed UAMPSBL-PCI algorithm still outperforms other algorithms in terms of estimation accuracy.

Fig. 8 shows the channel estimation performance against the number of units $N$ on RIS. As the size of RIS increases from $9\times9$ to $27\times27$, the performance of all estimation algorithms degrades. From the perspective of channel matrices, the increment of $N$ implies the number of elements estimated by each algorithm increases, leading to a significant accumulation of errors. However, simulation results demonstrate that the proposed UAMPSBL-PCI algorithm outperforms other algorithms even under these more challenging conditions.

Fig. 9 shows the channel estimation performance against the number of UPA antennas $M$ at BS. As the number of antennas increases from $3\times3$ to $21\times21$, the estimation performance of all algorithms shows a trend of first increasing and then stabilizing. However, the lower bound reveals that the change in $M$ will theoretically have no effect on the estimation performance of the algorithm. The main reason for this trend is the sparsity change of the estimated cascaded channel. Since the sparsity of the cascaded channel is low when $M$ is small, estimation algorithms based on the sparse recovery theory are unable to fully exploit their advantages, and as a result, perform poorly. With the increase of $M$, the cascaded channels become more sparse, and the performance of each algorithm is also improved continuously and approaches the theoretical limit gradually. In comparison, the proposed UAMPSBL-PCI algorithm still exhibits good performance when $M$ is small, proving its wide applicability.

\begin{figure}[!t]
\centering
\includegraphics[width=3.5in]{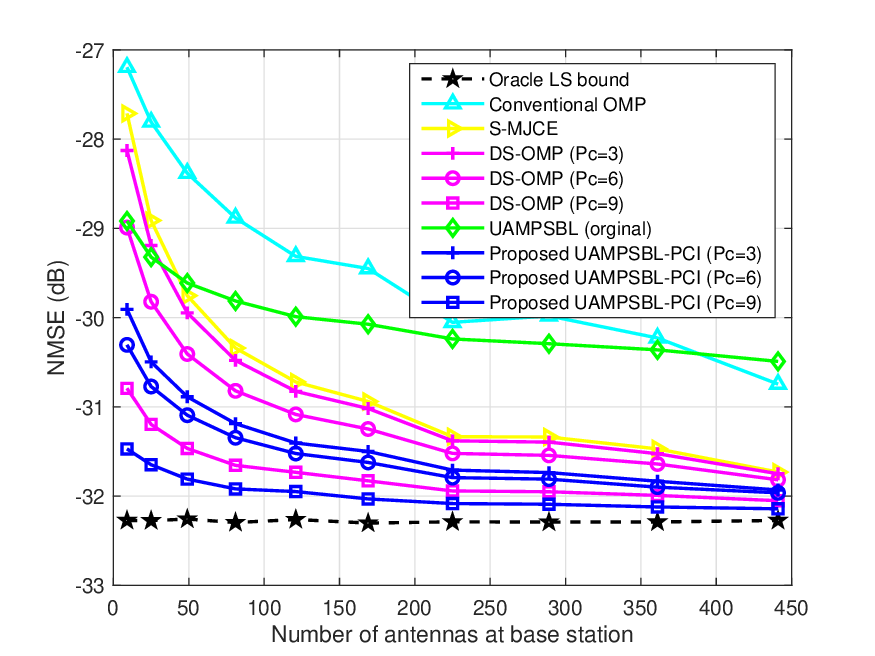}
\caption{Estimation performance NMSE against the number of antennas $M$ at BS in Scenario 1. }
\label{fig_9}
\end{figure}

\begin{figure}[!t]
\centering
\includegraphics[width=3.5in]{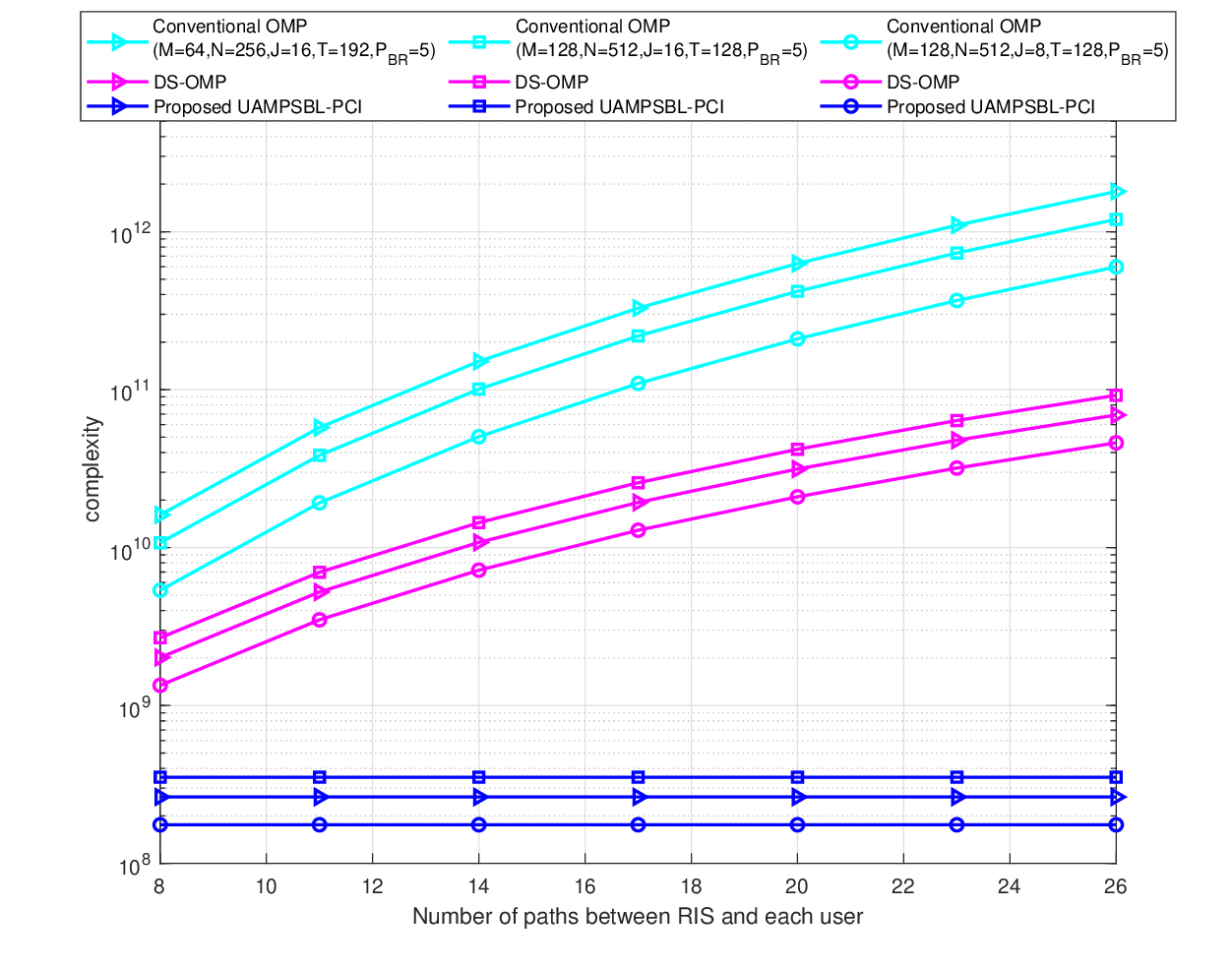}
\caption{Complexity comparison against the number of paths between RIS and each user.}
\label{fig_10}
\end{figure}

As shown in Fig. 10, we conducted a complexity comparison of the proposed UAMPSBL-PCI, DS-OMP and the conventional OMP algorithm with respect to the number of paths $P_{j}$ between the RIS and each user. As mentioned earlier, the proposed UAMPSBL-PCI algorithm has a complexity of $\mathcal{O}(JTM)+\mathcal{O}(JTN{P}_{BR}I)$, while the conventional OMP algorithm has a complexity of $\mathcal{O}(JTN{P}_{BR}{{P}_{j}}^4)$ \cite{RN25}, and the DS-OMP algorithm has a complexity of $\mathcal{O}(JTM)+\mathcal{O}(JTN{P}_{BR}{{P}_{j}}^3)$ \cite{RN10}. Numerous experiments suggest that the number of estimation iterations, denoted as $I$, in algorithm UAMPSBL-PCI does not exceed 65. It is evident that the proposed UAMPSBL-PCI algorithm has a lower complexity under the same system settings and is not affected by the number of paths $P_{j}$, which is a crucial factor in selecting a channel estimation algorithm for the RIS-assisted communication system.

\begin{figure}[!t]
\centering
\includegraphics[width=3.5in]{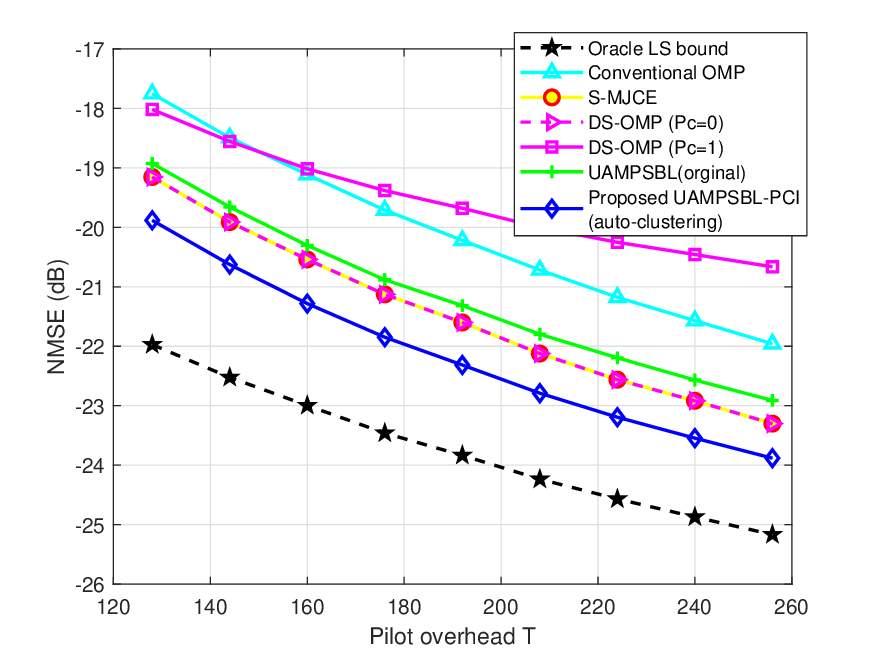}
\caption{Estimation performance NMSE against the pilot overhead $T$ in Scenario 2.}
\label{fig_11}
\end{figure}

\begin{figure}[!t]
\centering
\includegraphics[width=3.5in]{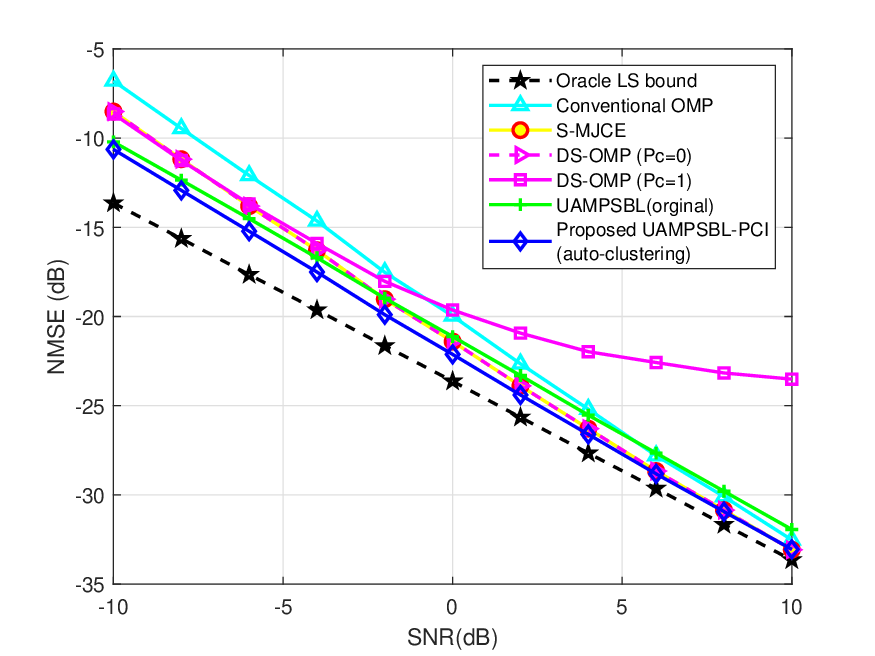}
\caption{Estimation performance NMSE against the SNR in Scenario 2.}
\label{fig_12}
\end{figure}

\subsection{Simulation results for Scenario 2}
The channel estimation performance of the proposed UAMPSBL-PCI algorithm for the RIS-assisted communication system in the more general scenario where subsets of users share common channels is presented in this subsection.

Considering that there is no common column support shared by all users in this more general system, $P_{c}$ will be set to 0. In order to explore the impact of the common column $P_{c}$ on the DS-OMP algorithm in this scenario, we also employ the DS-OMP $(Pc=1)$ algorithm as a control group. Assuming that $J$ users are randomly divided into three clusters, users in the same cluster have random $v, v\leq P_{j}$ common columns, while users in different clusters may or may not have common column support. Under this system setting, the performance of each estimation algorithm is evaluated.

Fig. 11 shows the channel estimation performance against the pilot overhead $T$. Similar to the simulation results in the previous subsection, the performance of all channel estimation algorithms improves with increasing pilot overhead. Since there is no common column shared by all users at this time, the DS-OMP $(P_{c}=1)$ algorithm misuses the column support, which leads to a significant drop in its performance. At this point, the DS-OMP algorithm that only considers row support $(P_{c}=0)$ has a similar performance to the S-MJCE algorithm. The proposed UAMPSBL-PCI algorithm with an auto-clustering mechanism requires the least pilot overhead to obtain the same estimation accuracy.

Fig. 12 shows the channel estimation performance against SNR. As expected, the estimation accuracy of all algorithms improves with increasing SNR. Notably, the proposed UAMPSBL-PCI algorithm consistently exhibits better performance compared to other investigated algorithms. Its superiority is particularly evident under unfavorable channel conditions. Furthermore, it is worth noting that the DS-OMP algorithm with column support is no longer applicable.

\begin{figure}[!t]
\centering
\includegraphics[width=3.5in]{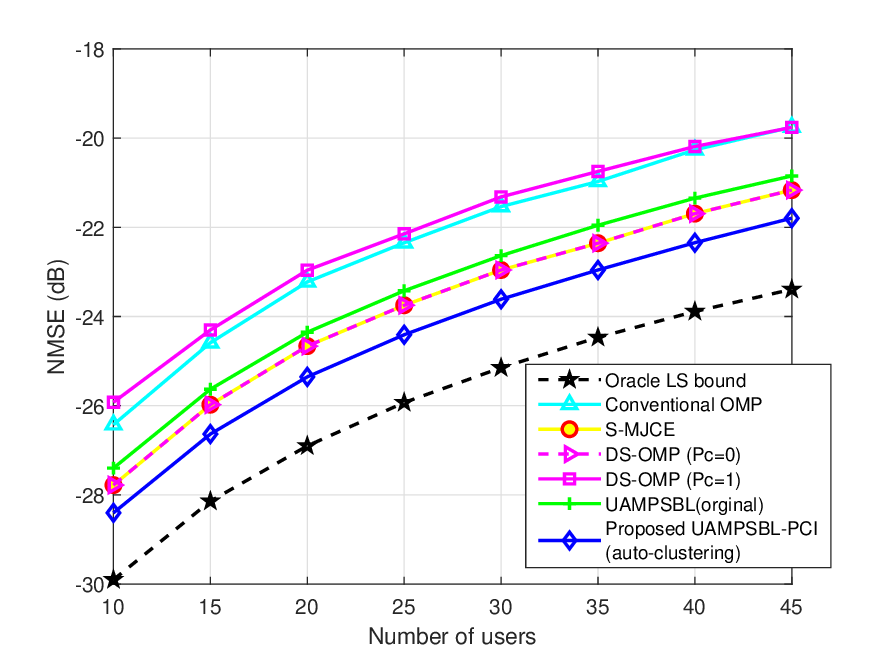}
\caption{Estimation performance NMSE against the number of users $J$ in Scenario 2.}
\label{fig_13}
\end{figure}

\begin{figure}[!t]
\centering
\includegraphics[width=3.5in]{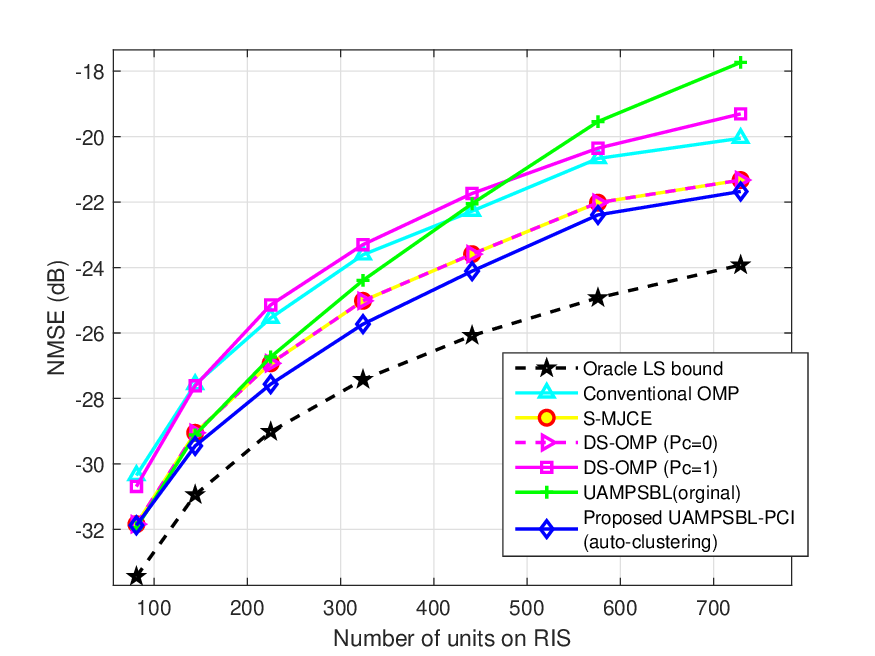}
\caption{Estimation performance NMSE against the number of units  $N$ on RIS in Scenario 2. }
\label{fig_14}
\end{figure}

Fig. 13 and 14 illustrate the performance of these channel estimation algorithms as the number of users $J$ and the number of units $N$ on the RIS increase, respectively. Both simulation results exhibit a similar trend to the simple system, where the estimation performance decreases as the corresponding parameter increases. However, the proposed UAMPSBL-PCI algorithm still outperforms the others in comparison.

Fig. 15 shows the channel estimation performance against the number of UPA antennas $M$ at BS. As the sparsity of the channel increases with $M$, the possibility of common column support between all users decreases, leading to a sharp decline in the performance of the DS-OMP $(Pc=1)$ algorithm, which misuses column support. Meanwhile, other algorithms continue to show an increasing-then-stabilizing trend, and the proposed UAMPSBL-PCI algorithm continues to demonstrate the best performance.

\begin{figure}[!t]
\centering
\includegraphics[width=3.5in]{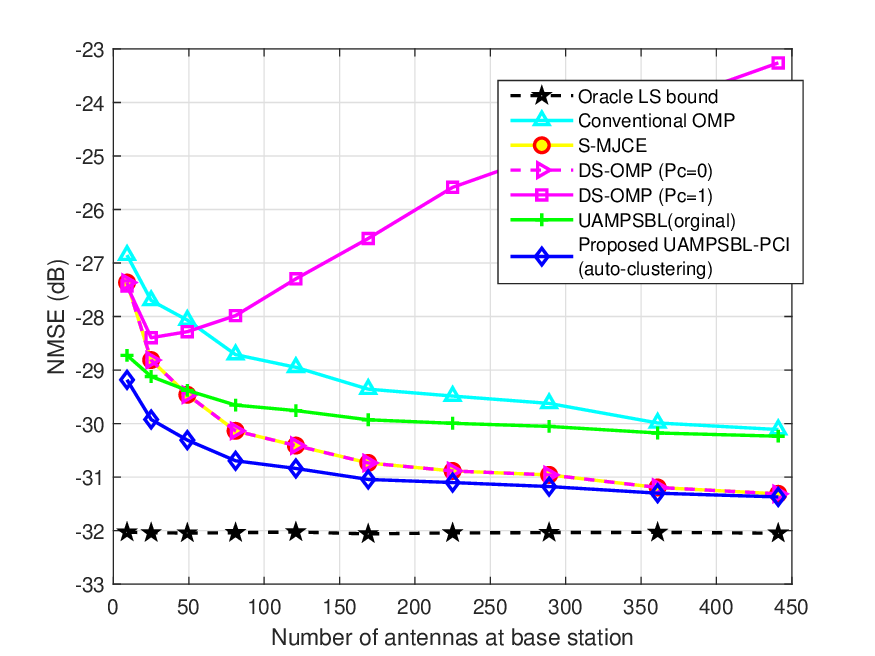}
\caption{Estimation performance NMSE against the number of antennas $M$ at BS in Scenario 2. }
\label{fig_15}
\end{figure}

\section{Conclusion}
In this paper, we investigated the cascaded channel estimation problem for a fully passive single RIS-assisted communication system. We first explored the dual-structure sparsity of the angular cascaded channel and extended it from a simple scenario where all users share the common channels to a more general scenario where subsets of users share the common channels. To make full use of the sparse nature of the channel, we proposed a new channel estimation algorithm called UAMPSBL-PCI based on the plain UAMP-SBL algorithm. Specifically, the algorithm employs a fast scan mechanism to obtain sparse row-column support as prior information. This prior information is incorporated into the subsequent estimation process, thereby enhancing the accuracy of the channel information. Additionally, we proposed an automatic clustering mechanism to extend it to more general scenarios. Numerous simulation experiments demonstrated the superiority of the proposed UAMPSBL-PCI algorithm over existing channel estimation methods in terms of accuracy, complexity, and wide applicability. In subsequent work, we will examine the application potential of the proposed algorithm in a terahertz channel  and attempt to enhance its performance through neural network processing.

% \bibliographystyle{IEEEtran}
% \bibliography{ref.bib}


\end{document}